\documentclass[a4paper,prd,superscriptaddress,11pt,noshowkeys,notitlepage,nofootinbib]{revtex4}
\usepackage[utf8]{inputenc}
\usepackage{ae,aecompl}
\usepackage{amsmath,amssymb,mathrsfs}
\usepackage{graphicx}
\usepackage{ccnishi}
\usepackage{dsfont}
\usepackage{slashed}
\usepackage{units}
\usepackage{nicefrac}

\usepackage[usenames,dvipsnames]{xcolor} % svgnames
\usepackage{hyperref}
\hypersetup{
    colorlinks=true,        % false: boxed links; true: colored links
    linkcolor=purple,       % color of internal links, e.g., eqs.
    citecolor=JungleGreen,  % color of links to bibliography
    filecolor=magenta,      % color of file links
}

\providecommand{\nhalf}{\nicefrac{1}{2}}

\providecommand{\cM}{\mathscr{M}}

\providecommand{\cY}{\mathscr{Y}}
\providecommand{\cO}{\mathcal{O}}

\providecommand{\dst}{\displaystyle}
\providecommand{\tY}{\tilde{Y}}

\providecommand{\ckmsm}{V_{\rm ckm}^{\rm sm}}

\providecommand{\lag}{\mathscr{L}}

\providecommand{\eps}{\epsilon}

\providecommand{\multi}{\textsc{MultiNest}}
\providecommand{\pymulti}{\textsc{PyMultiNest}}

\makeatletter
\newcommand*\bigcdot{\mathpalette\bigcdot@{.5}}
\newcommand*\bigcdot@[2]{\mathbin{\vcenter{\hbox{\scalebox{#2}{$\m@th#1\bullet$}}}}}
\makeatother

\begin{document}
%%%%%%%%%%%%%%%%%%%%%%%%%%%%%%%%%%%%%%%%%%%%%%%%%
\title{
Flavor constraints for a vector-like quark of Nelson-Barr type
}
\author{A.~L.~Cherchiglia}
\thanks{E-mail: adriano.cherchiglia@ufabc.edu.br}
\affiliation{Centro de Ci\^{e}ncias Naturais e Humanas,
Universidade Federal do ABC, 09.210-170,
Santo Andr\'{e}-SP, Brasil}
\author{G.~De~Conto}
\thanks{E-mail: george.de.conto@gmail.com}
\affiliation{Centro de Ci\^{e}ncias Naturais e Humanas,
Universidade Federal do ABC, 09.210-170,
Santo Andr\'{e}-SP, Brasil}
\author{C.~C.~Nishi}
\thanks{E-mail: celso.nishi@ufabc.edu.br}
\affiliation{Centro de Matem\'{a}tica, Computa\c{c}\~{a}o e Cogni\c{c}\~{a}o,
Universidade Federal do ABC, 09.210-170,
Santo Andr\'{e}-SP, Brasil}
%\affiliation{Universidade Federal do ABC - UFABC, Santo Andr\'{e}, SP, Brasil}

% \date{\today}
%%%%%%%%%%%%%%%%%%%%%%%%%%%%%%%%%%%%%%%%%%%%%%%%%
\begin{abstract}
The Nelson-Barr (NB) mechanism to solve the strong CP problem assumes CP conservation, arranges vanishing $\bar{\theta}$ at tree-level and requires vector-like quarks (VLQs) to transmit the CP breaking to the SM.
We analyze the flavor constraints coming from the presence of one such down type VLQ of NB type by performing a global fit on the relevant flavor observables. A comparison is made to the case of one generic VLQ. We find that the allowed parameter space for the VLQ Yukawa couplings and the mixing to the SM are confined to a region much smaller than in the generic case, making the NB case falsifiable in principle.
\end{abstract}
%%%%%%%%%%%%%%%%%%%%%%%%%%%%%%%%%%%%%%%%%%%%%%%%%
% \pacs{12.60.Fr, 14.80.Cp, 11.30.Qc, 02.20.-a}
%\keywords{ }
%\twocolumn
\maketitle
% \tableofcontents
%%%%%%%%%%%%%%%%%%%%%%%%%%%%%%%%%%%%%%%%%%%
\section{Introduction}
\label{sec:intro}

The Nelson-Barr (NB) solution\,\cite{nelson,barr} to the strong CP problem is a very simple mechanism based on the assumption that CP is a fundamental symmetry of nature\,\cite{strongCP:CP}\footnote{%
Alternatively, one can assume P is a fundamental symmetry\,\cite{strongCP:P}. We will focus on the solution based on CP.
}
---hence the QCD vacuum angle is zero at tree level--- spontaneously broken so that the CP violation we observe in the SM is reproduced. The mechanism proposes a simple recipe to arrange a vanishing contribution to $\bar{\theta}$ at tree level from quark Yukawa couplings, but still allows large CP violation to arise in the SM.
Then the challenge\,\cite{dine,perez.shalit} is to keep the calculable radiative corrections to $\bar{\theta}$ to be tiny, $\bar{\theta}\lesssim 10^{-10}$, conforming to the experimental observations\,\cite{EDMreview,nEDM:exp,EDM:Hg}.

A necessary byproduct of the NB mechanism is the presence of vector-like quarks (VLQ) whose mixing with ordinary quarks after spontaneous CP breaking is the only source of CP breaking that is transmitted to the SM already at tree level.\,\footnote{%
Large CKM CP violation may be also generated at loop level, although it requires strong coupling in the mixing term and supersymmetry to protect $\bar{\theta}$\,\cite{hiller:01}.
}
A simple definition of these VLQs of Nelson-Barr type (NB-VLQs) was given recently for singlet VLQs in Ref.\,\cite{nb-vlq}.%
\footnote{A similar setting was proposed some time ago in Ref.\,\cite{lavoura}.}
The simplest complete model was proposed 30 years ago by Bento, Branco and Parada\,\cite{BBP}, which contained one down-type VLQ and one complex scalar responsible for CP breaking.
More recently, Ref.\,\cite{NB:CP4} improved on this model by imposing a non-conventional CP symmetry which required the addition of two VLQs as the price to protect $\bar{\theta}$ from receiving one-loop corrections.
Similarly to these UV completions, the implementation of the NB idea generically leads to theories where CP violation arises only in wave-function renormalization\,\cite{hiller:01}.
Flavor alignment may follow if one combines the spontaneous breaking of CP with the breaking of quark family number\,\cite{meade:SFV}.
See other approaches in Refs.\,\cite{scpv:others:recent,cp-texture}.

However, VLQs appear as byproducts in many other theories, many unrelated to the strong CP problem.
Little higgs\,\cite{little.higgs} and composite Higgs\,\cite{composite.higgs} theories comprise a small subset.\footnote{%
Ref.\,\cite{perez.shalit} improves the quality of the NB solution by making the spontaneous CP breaking sector composite.
}
Therefore, if VLQs are ever found at the TeV scale, it is pressing that we explore the characteristic features of VLQs related to the strong CP problem and the origin of CP violation in nature.

Here we study in more detail the case of the presence of a NB-VLQ of charge $-1/3$, focusing on the flavor constraints. We perform a global fit on a set of flavor observables and compare the effect of a NB-VLQ with a VLQ of generic type, not necessarily related to the source of CP violation.
For one down-type NB-VLQ, Ref.\,\cite{nb-vlq} has shown that the VLQ couplings to the up type quarks through the charged current necessarily follow a hierarchical pattern, a feature that suppresses the flavor changing effects involving lighter quarks. 
This nongeneric and hierarchical pattern is interesting if we consider the new physics scales that we can probe 
by, e..g, $\Delta F=2$ transitions induced by new physics contributions to flavor changing operators of the form 
\eq{
\label{Lambda.scale}
\frac{1}{\Lambda_{ij}^2}(\bar{d}_{iL}\gamma^\mu d_{j_L})^2\,.
}
A generic analysis\,\cite{isidori.cern.12} at each sector $(ij)=(sd),(bd),(bs)$ leads to the rough bound%
\footnote{More precise but still generic bounds can be found in the same references.}
\eq{
\frac{1}{\Lambda_{ij}} \lesssim \frac{|V_{ti}^*V_{tj}|}{3.4\,\unit{TeV}}\,,
}
which follows the hierarchy
\eq{
\Lambda_{sd}^{-1}:\Lambda_{bd}^{-1}:\Lambda_{bs}^{-1}\sim 0.009:0.2:1\,.
}
For one NB-VLQ, this kind of hierarchy is automatically induced.\footnote{%
The overall scale, however, is not determined.
}
We will obtain quantitative bounds from the global fit.

We organize this paper as follows: in Sec.\,\ref{sec:model} we briefly review the model of one NB-VLQ and describe the main features. This includes the \emph{seesaw parametrization} that was developed previously\,\cite{nb-vlq}.
In Sec.\,\ref{sec:pheno} we list the observables that will be used in the global fit.
The methods and the results of the global fit are shown in Sec.\,\ref{sec:num}.
We finally conclude in Sec.\,\ref{sec:concl}.

%%%%%%%%%%%%%%%%%%%%%%%%%%%%%%%%%%%%%%%%%%%
\section{Review of the model and parametrization}
\label{sec:model}

Our model of one down-type VLQ of Nelson-Barr type (NB-VLQ), denoted by $B_{L,R}$, is defined by the Lagrangian\,\cite{nb-vlq}
\eqali{
\label{yuk:NB}
-\lag&=\bar{q}_{iL}\cY^d_{ij} Hd_{jR}+\bar{q}_{iL}\cY^u_{ij} \tilde{H}u_{jR}
\cr
&\quad +
\bar{B}_{L}\cM^{Bd}_{j} d_{jR}+\bar{B}_{L}\cM_B B_{R}+h.c.,
}
where $i,j=1,2,3,$ and 
with the additional requirement that $\cY^u,\cY^d$ are \textit{real} $3\times 3$ matrices, $\cM_B$ is a real mass and only $\cM^{Bd}$ is a complex row vector. 
This structure follows from CP conservation and a $\ZZ_2$ symmetry\,\footnote{%
A larger $\ZZ_n$ or $U(1)$\,\cite{dine}, a nonabelian global or gauge symmetry\,\cite{nelson,barr} can be also used.
Note that the definition based on $\ZZ_2$ is not sufficient when singlet and doublet VLQs are simultaneously present. 
}
under which only $B_{L,R}$ are odd, and only $\cM^{Bd}$ breaks CP and $\ZZ_2$ softly (spontaneously) realizing the Nelson-Barr mechanism that guarantees $\bar{\theta}=0$ at tree-level\,\cite{nelson,barr}.
The SM fields are $q_{iL}, d_{iR}$ and $u_{iR}$, corresponding to the quark doublets, down-type singlets and up-type singlets, respectively.
This Lagrangian contains $3~(\cY^u)+9~(\cY^d)+2~(\cM^{Bd})+1~(\cM_B)=15$ parameters in total (see explicit parametrization below) using the freedom of \textit{real} orthogonal transformations in the $\cY^u$ diagonal basis\,\cite{nb-vlq}.

In contrast, a \textit{generic} VLQ is customarily described by the Lagrangian 
\eqali{
\label{yuk:VLQ}
-\lag&=\bar{q}_{iL}Y^d_{ij} Hd_{jR}+\bar{q}_{iL}Y^u_{ij} \tilde{H}u_{jR}
\cr &~~+\ 
\bar{q}_{iL} Y^B_{i}H\,B_{R} 
+ \bar{B}_{L}M_B B_{R}+h.c.,
}
where $M_B$ is expected to be much larger than the electroweak scale.
If generic, this lagrangian depends on $3~(Y^u)+7~(Y^d)+5~(Y^B)+1~(M_B)=16$ parameters using now the freedom of unitary rotations in $Y^u$ diagonal basis.
We can see that one more parameter is needed compared to the case of one NB-VLQ. Hence, the NB case is just a subcase and when the lagrangian \eqref{yuk:NB} is rewritten in the form \eqref{yuk:VLQ}, the various parameters cannot be independent and correlations necessarily appear\,\cite{nb-vlq}.
In special, \emph{only one} CP violating parameter controls all CP violation in the NB case while the generic case depends on three CP violating parameters\,\cite{lavoura.branco,branco:book,nir:slac99}.
We emphasize this information in table \ref{tab.1}.
\begin{table}[h]
\[
\begin{array}{|c|cc|}
\hline
                    & \text{\# of param.} & \text{\# of CP odd}\cr
\hline
\text{SM}           & 10                & 1                \cr
\text{One generic VLQ}  & 16                 & 3         \cr
\text{One NB-VLQ}       & 15                 & 1        \cr
\hline
\end{array}
\]
\caption{\label{tab.1}
Number of parameters in the flavor sector of the SM and with the addition of one VLQ.
}
\end{table}

The changing of basis from \eqref{yuk:NB} to \eqref{yuk:VLQ} is easily described by comparing the $4\times 4$ mass matrix of the down-type quarks following from \eqref{yuk:NB} and \eqref{yuk:VLQ}, respectively, after EWSB:
\eq{
\label{mass.matrix}
\text{NB:}\quad\cM^{d+B}=\mtrx{\frac{v}{\sqrt{2}}\cY^d & 0\cr \cM^{Bd}  & \cM_B}
\,,\quad
\text{generic:}\quad M^{d+B}=\mtrx{\frac{v}{\sqrt{2}}Y^d & \frac{v}{\sqrt{2}}Y^B\cr 0 & M_B}\,.
}
Only a unitary transformation from the right is necessary to connect them:
\eq{
\cM^{d+B}W_R=M^{d+B}\,.
}
The relevant exact relations are given by
\subeqali{
\label{MB}
M_B&=\sqrt{\cM^{Bd}{\cM^{Bd}}^\dag+\cM_B^2}\,,
\\
\label{YB:NB}
Y^B &=\cY^d {\cM^{Bd}}^\dag/M_B\,,
\\
\label{Yd:NB}
Y^d{Y^d}^\dag &=\cY^d\left(\id_3-{\cM^{Bd}}^\dag\cM^{Bd}/M_B^2\right){\cY^d}^\tp\,.
}
For both NB and generic cases, the form $M^{d+B}$ has a simple interpretation in the leading quark seesaw approximation\footnote{%
In this approximation, the block diagonalized form of $M^{d+B}$ is simply given by neglecting the $Y^B$ term off the diagonal. At this order, $Y^B$ will only contribute to the VLQ mixing with the SM but not to the masses.
} 
where $M_B\gg v$:
$M_B$ is the VLQ mass while $Y^d{Y^d}^\dag$ is the leading SM down-type Yukawa matrix squared which contains the CKM CP violation\,\cite{BBP}.
Therefore, in leading order, the latter is determined from SM input:
\eq{
\label{Yd:sm.input}
Y^d{Y^d}^\dag =\frac{2}{v^2}V_{d_L}\diag(m^2_d,m^2_s,m^2_b)V_{d_L}^\dag\,,
}
where $V_{d_L}$ is the CKM matrix $\ckmsm$, except for possible phases from the left.
If $V_{d_L}=\ckmsm$, note that the righthand side of \eqref{Yd:sm.input} contains 7 parameters from the SM down sector.

Removing from the counting the three up quark Yukawa couplings in the diagonal basis, we can now concentrate on the 12 parameters in $\cM^{d+B}$ for the NB-VLQ case or 13 parameters in $M^{d+B}$ for the generic case; cf.\,\eqref{mass.matrix}.
We should now concentrate on how to separate the 7 SM input parameters from the BSM parameters.

In the generic case, $Y^d$ and $Y^B$ are independent. Then, in the seesaw approximation, 7 parameters in $Y^d$ can be fixed from \eqref{Yd:sm.input}, with $V_{d_L}=\ckmsm$, and the remaining 5 parameters in $Y^B$ and one parameter in $M_B$ are free BSM parameters.

In the NB case, $Y^d$ and $Y^B$ are \emph{not} independent and a different parametrization is needed.
We now describe the seesaw parametrization devised in Ref.\,\cite{nb-vlq} for one NB-VLQ.

We first choose \eqref{MB} as one of the parameters instead of $\cM_B$ and rewrite
\eq{
\cM^{d+B}=\mtrx{\frac{v}{\sqrt{2}}\cY^d & 0\cr M_B w^\dag & M_B\sqrt{1-|w|^2}}\,.
}
Without loss of generality\,\cite{nb-vlq},
\eq{
\label{w}
w={\cM^{Bd}}^\dag/M_B=\mtrx{0\cr ib \cr a}\,,
}
with $b<a$ and $a^2+b^2< 1$.
So $\{M_B,w\}$ contains three parameters while $\cY^d$, being real and generic, contains 9 parameters.

The combination in the righthand side of \eqref{Yd:NB} depends on $\cY^d$ and $w$, and is partially fixed by SM input in \eqref{Yd:sm.input}.
However, as is clear from Eqs.\,\eqref{YB:NB} and \eqref{Yd:NB}, both $Y^d$ and $Y^B$ depends on $\cY^d$ and $w$, and we cannot transfer phases from one to another.
We then need to consider additional phases\footnote{%
In Ref.\,\cite{nb-vlq}, these phases were denoted as $\beta_2,\beta_3$, respectively.} 
in
\eq{
\label{VdL}
V_{d_L}=\mtrx{1&&\cr&e^{i\beta_1}&\cr&&e^{i\beta_2}}V^{\rm sm}_{\rm ckm}\,,
}
in order to take into account SM input in \eqref{Yd:sm.input}.

Given $\beta_1,\beta_2$, the real matrix $\cY^d$ is partially determined from the inversion formula for \eqref{Yd:NB}; see appendix \ref{ap:inversion} and Ref.\,\cite{nb-vlq} for details.
The remaining freedom is parametrized by $b$ in $w$ and one additional angle parameter $\gamma$.
In this way, in our seesaw parametrization, among the 12 parameters in $\cM^{d+B}$, the 7 parameters are fixed from the SM input while the following 5 BSM parameters are free:
\eq{
\label{param.bsm}
\{M_B,b,\gamma,\beta_1,\beta_2\}\,.
}
So $\cY^d$ and $w$ depend on $b,\gamma,\beta_1,\beta_2$, and $M_B$ can be chosen independently.
In special, we restrict
\eq{
\label{b.range}
b\in [0.02,1/\sqrt{2}],
}
so that the seesaw approximation is valid within 1\% for the mass eigenvalues and the moduli of the $3\times 3$ block of the CKM matrix\,\cite{nb-vlq}. The angle $\gamma$ and phases $\beta_1,\beta_2$ all vary in the whole range $[0,2\pi]$.

In the generic as well as NB case, the connection with observables are all encoded in the rectangular CKM matrix $V\sim 3\times 4$, 
\eq{
\label{V}
V=U_{u_L}^\dag P U_{d_L}\approx
% \left(\begin{array}{c|c}
% (\id_3-\ums{2}\theta_L\theta_L^\dag)V_{d_L} & \theta_L V_{B_L}
% \end{array}\right)\,.
V_{d_L}
\left(\begin{array}{c|c}
\id_3-\ums{2}\delta X^d & \Theta
\end{array}\right)\,,
}
where $U_{d_L}$ is the matrix that diagonalizes either matrix in \eqref{mass.matrix} from the left, 
$P$ is a $3\times 4$ projection matrix, nonvanishing only for $P_{11}=P_{22}=P_{33}=1$, and
\eq{
\label{Theta}
\Theta\equiv V_{d_L}^\dag \theta_L\,,
\quad
\delta X^d\equiv \Theta\Theta^\dag\,.
}
The diagonalizing matrix of the up sector can be ignored, $U_{u_L}=\id_3$, in the basis where $Y^u=\hat{Y}^u$.
The expansion parameter in the seesaw is given by
\eq{
\label{thetaL}
\theta_L=\frac{v}{\sqrt{2}}Y^B M_B^{-1}\,.
}
The square matrix $X^d$ of size $4$ that describes the FCNC coupling to $Z$ is
\eq{
\label{X:d}
X^d=V^\dag V
\approx 
% \left(\begin{array}{c|c}
% V_{d_L}^\dag(\id_3-\ums{2}\theta_L\theta_L^\dag)^2V_{d_L}  & * 
% \cr\hline
% V_{B_L}^\dag\theta_L^\dag(\id_3-\ums{2}\theta_L\theta_L^\dag)V_{d_L}  & V_{B_L}^\dag\theta_L^\dag\theta_LV_{B_L} 
% \end{array}\right)\,.
\left(\begin{array}{c|c}
\id_3-\delta X^d & \Theta
\cr\hline
*  & \Theta^\dag\Theta
\end{array}\right)\,.
}
So we see that all couplings of VLQs to gauge bosons depend solely on the matrix $\Theta$ within the seesaw approximation.

For definiteness, we employ the phase convention for $V$ such that
\eq{
\label{BLS}
\arg(V)=\mtrx{0&0& * & *\cr \pi & * & 0 & *\cr *&*& 0& 0}\,.
}
The stars denote free phases.
This convention exhausts all the rephasing freedom and it is easier to implement than the one coming from the angle-phase parametrizations.
For the $3\times 3$ block, this convention is similar to the one employed in Ref.\,\cite{branco:book} and differs slightly only by a tiny phase in the second column.
This convention is also very close to the standard parametrization of the $3\times 3$ CKM or to the angle-phase parametrization of the $4\times 4$ unitary matrix employed in Ref.\,\cite{nb-vlq}. They differ by tiny phases once SM input is considered.

The phase convention \eqref{BLS} can be chosen directly for the generic case but it is not automatic for the NB case as the diagonalizing matrix for \eqref{mass.matrix} has implicit phase relations.
We enforce it after the diagonalization process, a procedure which also redefines $Y^d$ and $Y^B$ by rephasing from the left.

In Ref.\,\cite{nb-vlq}, it was shown that once the SM input was considered, the mixing of the heavy VLQ with the SM quarks followed a hierarchical structure:
\eq{
\label{ViB:hierarchy}
|V_{uB}|\ll |V_{cB}|\ll |V_{tB}|\,.
}
For a parameter $b$ not so small, the hierarchy roughly follows the hierarchy of $b$-quark mixing:
\eq{
\label{ViB:3f}
|V_{uB}|:|V_{cB}|:|V_{tB}|\sim |V_{ub}|:|V_{cb}|:|V_{tb}|\approx 0.004:0.04:1\,,
}
a feature that is exact in case the VLQ couples exclusively with the third family.
The hierarchy also applies to $Y^B_i$ since they are proportional to 
$V_{iB}$ at leading order, cf.\,\eqref{thetaL}.\footnote{%
See appendix \ref{ap:approx} for an approximate function describing the upper boundary of $|Y^B_i|$ as a function of $b$.
}
This hierarchy largely renders the model flavor safe as the most restrictive flavor constraints of flavor changing among the first and second families are naturally suppressed.
Here we will test these constraints more quantitatively by performing a global fit.

If BSM models are studied in terms of effective field theories and matched onto the SM or effective theories at lower energies, flavor observables can be written in terms of SM parameters and the BSM parameters $Y^B_i$ and the VLQ mass $M_B$.
Constraints can then be placed on these BSM parameters.
This study was performed in, e.g., Ref.\,\cite{buras.celis.17}.
Information on $Y^B$ is equivalent to the information on the CKM matrix $V$ and at leading order $Y^B$ is proportional to the light-heavy mixing parameters $\theta_L$ in \eqref{thetaL}.
We can extract $Y^B$ independently of the basis for right-handed fields in terms of the mass matrix 
in \eqref{mass.matrix} as
\eq{
\frac{v}{\sqrt{2}}Y^B_i=\frac{(M^{d+B}{M^{d+B}}^\dag)_{i4}}{\sqrt{(M^{d+B}{M^{d+B}}^\dag)_{44}}}\,.
}
This relation is also valid if we replace $M^{d+B}$ by $\cM^{d+B}$ from the NB case since the two matrices are related by unitary transformation for right-handed quarks.
For the generic case, the mass matrix squared $(M^{d+B}{M^{d+B}}^\dag)$ can be reconstructed from the $4\times 4$ mixing and masses\,\footnote{%
Note that the \emph{mass} $M_B$ here only coincides with the \emph{parameter} $M_B$ in \eqref{mass.matrix} in the leading seesaw approximation.
} as
\eq{
M^{d+B}{M^{d+B}}^\dag=U_{d_L}\diag(m_d^2,m_s^2,m_b^2,M_B^2)U_{d_L}^\dag\,,
}
while for the NB case $\cM^{d+B}$ itself can be used.
The $4\times 4$ diagonalizing matrix $U_{d_L}$ can be parametrized in terms of the standard parametrization for the $3\times 3$ CKM matrix and additional angles, $\theta_{14},\theta_{24},\theta_{34}$, and additional phases $\delta_1,\delta_2$\,\cite{nb-vlq}; see also Refs.\,\cite{gouvea,botella}.
When a specific basis is understood, $Y^B$ will denote the Yukawa coupling in the basis where the up-type Yukawas are diagonal $Y^u=\hat{Y}^u$.
However, to compare with Ref.\,\cite{buras.celis.17}, we will also need
\eq{
\label{def:YB-tilde}
\tY^B_i\equiv (V^\dag_{d_L}Y^B)_i\,,
}
defined in the weak basis where $Y^d$ is diagonal when $Y^B$ is defined in the basis where $Y^u$ is diagonal.

%%%%%%%%%%%%%%%%%%%%%%%%%%%%%%%%%%%%%%%%%%%
\section{Phenomenological constraints for $n_B=1$}
\label{sec:pheno}

In this section we describe the various observables that will be considered in the fit.
Other observables will be predicted from the fit and these are discussed as well.

\subsection{VLQ mass}

Because $V_{iB}$ are hierarchical, the $B$ quark couples dominantly with the top and we can use the current constraint coming from direct searches at the LHC\,\cite{atlas,cms}:
\eq{
\label{MB:limit}
M_B\gtrsim 1.2\,\unit{TeV}\,.
}
Considering the recent analysis \cite{cms:20}, we will perform our fit for the conservative value $M_B=1.4\,\unit{TeV}$.

Even if $V_{iB}$ are hierarchical and flavor alignment is naturally enforced, we need to test the structure of $V$ against the various flavor observables.
The various observables that we will use to constrain the model will be described in the following subsections and is summarized in Table \ref{tab:obs}.

We reserve the letter $B$ for the heavy quark and the $B$-mesons will be always denoted with their subscripts $B_d^0$ and $B_s^0$.

\subsection{Constraints from $|V_{ij}|$}

Most of the moduli $|V_{ij}|$ of the CKM matrix are extracted from tree level processes.
The exceptions are $|V_{td}|$ and $|V_{ts}|$ which are extracted from $B_d^0$ and $B_s^0$ meson oscillations through box diagrams involving the top quark.
We impose direct constraints on the ones extracted at tree level while the constraints for $|V_{td}|$ and $|V_{ts}|$ are extracted directly from $\Delta m_{B_d}$ and $\Delta m_{B_s}$ by considering the VLQ contribution; see Sec.\,\ref{sec:Delta.mB}. So these entries may deviate from the SM values.

The experimental values for $|V_{ij}|$ are\,\cite{PDG}:
\eq{
\label{|V|:exp}
|V_{ij}|_{\rm exp}=
\mtrx{
0.97370(14) & 0.2245(8) & 3.82(0.24)\times 10^{-3}
\cr
0.221(4) & 0.987(11) & 41.0(1.4)\times 10^{-3}
\cr
\underline{8.0(0.3)\times 10^{-3}} & \underline{38.8(1.1)\times 10^{-3}} & 1.013(30)
}\,.
}
The values for $|V_{td}|$ and $|V_{ts}|$, which are underlined, will not be considered directly.
Note that \eqref{|V|:exp} shows a somewhat lower value for $|V_{ud}|$\,\cite{PDG} due to use of updated hadronic uncertainty\,\cite{ramsey:Vud}.
These values hint at nonunitarity of the first row of the CKM matrix which can be interpreted in terms of VLQs\,\cite{berezhiani.19,cheung.20}.

For comparison, we can also show the values for the magnitudes of the CKM elements obtained from the combination of the various experiments and assuming unitarity.
The result reported by PDG 2020\,\cite{PDG} is
\eq{
\label{|V|:exp:fit} 
|V_{ij}|_{\rm exp}^{\rm fit}=
\mtrx{
0.97401(11) & 0.22650(48) & 0.00361^{+0.00011}_{-0.00009}
\cr
0.22636(48) & 0.97320(11) & 0.04053^{+0.00083}_{-0.00061}
\cr
\underline{0.00854^{+0.00023}_{-0.00016}} & \underline{0.03978^{+0.00082}_{-0.00060}} & 0.999172^{+0.000024}_{-0.000035}
}
\,.
}
The corresponding parameters in the standard parametrization read 
\eqali{
s_{12}&=0.22650(48)\,,&\quad  s_{13}&=0.00361^{+0.00011}_{-0.00009}\,,
\cr
s_{23}&=0.04053^{+0.00083}_{-0.00061}\,,&\quad \delta&=1.196^{+0.045}_{-0.043}\,,
}
while the Jarlskog invariant is $10^{5}{\times}J=3.00^{+0.15}_{-0.09}$.
In the Wolfenstein parametrization, the same fit leads to
\eqali{
\label{pdg:Alamb}
\lambda&=0.22650(48)\,,&\quad  A&=0.790^{+0.017}_{-0.012}\,,
\cr
\bar\rho&=0.141^{+0.016}_{-0.017}\,,&\quad \bar\eta&=0.357(11)\,.
}

\subsection{$\Delta m_{B_d}$ and $\Delta m_{B_s}$}
\label{sec:Delta.mB}

The CKM matrix entries $|V_{td}|$ and $|V_{ts}|$ in the SM are currently extracted from $B_d^0$ and $B_s^0$ meson oscillations through box diagram contributions.
In the presence of VLQs, additional contributions arise due to the exchange of the VLQs.
Or, in an EFT description, from the deviation from unitarity of the CKM matrix and FCNC through $Z$ exchange.

The $B^{0}_q-\bar{B}^{0}_q$ mixing for the $q=d,s$ mesons leads to the mass difference 
\eq{
\Delta m_{B_q} = 2 |M_{12}^{B_q}|\,,
}
as long-distance effects are negligible.
This quantity will be sensitive to 
\eq{
\lambda^t_{qb}\equiv V_{tq}^*V_{tb}\,.
}
The mixing in the presence of a VLQ can be calculated as\,\cite{saavedra:flavor}
\eq{
M_{12}^{B_q} = \frac{G_{F}^{2}m_{W}^{2}f^{2}_{B_q}\hat{B}_{B_q}m_{B_q^{0}}}{12\pi^{2}}\left[(\lambda^t_{qb})^2\eta^{B}_{tt}S_0(x_t)+\Delta_{B_q}\right]\,,
}
where $x_t\equiv m_t^2/m_W^2$ and
\eq{
\Delta_{B_q}=-8X_{qb}\lambda^t_{qb}\eta^B_{tt} Y_0(x_t)+\frac{4\pi s^2_w}{\alpha_e}\eta^{B}_Z X^2_{qb}
\,
}
is the VLQ contribution depending on $X_{qb}$.
We only retain the dominant top contribution in the box diagrams.
The dominant theoretical error comes from the combination $f_{B_q}\sqrt{B_{B_q}}$ of the meson decay constant and bag parameter.
Lattice calculations with three flavors report the average values $f_{B_d}\sqrt{B_{B_d}}=225(9)\,\unit{MeV}$ and $f_{B_s}\sqrt{B_{B_s}}=274(8)\,\unit{MeV}$\,\cite{flag.19}.
Other input parameters are shown in table \ref{tab:input}.
NLO QCD corrections from running from $m_W$ to $m_b$ is taken into account by the factors\,\cite{buras:qcd.eta} $\eta^{B}_{tt}=0.55$, $\eta^{B}_{Z}=0.57$, where we neglect the small errors. 
$S_0$ and $Y_0$ are Inami-Lim functions given in appendix \ref{ap:aux}.

\subsection{$Z\to bb$: $R_b$}

Given the mixing among the SM $d$-type quarks and the VLQ, not only FCNC will be generated but also diagonal couplings of SM $d$-type quarks to the $Z$ boson will be modified. These modifications can be strongly constrained by the ratio $R_{b}$\,\cite{saavedra:handbook}, since this observable shows a good agreement with the SM prediction; see Table \ref{tab:obs}.

For definiteness,  
\eq{
R_b=\frac{\Gamma(Z\to b\bar{b})}{\Gamma(Z\to\text{hadrons})}\,.
}
The possible deviations appear in the neutral current with d-type quarks
\eq{
\mathcal{L}_{Zqq} = -\frac{g}{2 c_{w}} \bar{q}\gamma^{\mu}\left(c_{Lq}P_{L} + c_{Rq}P_{R}\right)b Z_{\mu}\,,
}
where, in our model,
\eq{
c_{Lq} = c_{Lq}^{\rm SM} + \delta c_{Lq}, \quad c_{Lq}^{\rm SM} = - 1 + \frac{2}{3}s_w^{2}, \quad  \delta c_{Lq} = 1 - X^{d}_{qq}, \quad c_{Rq}=2/3s_w^{2}.
\label{eq:clq}
}

Since only small deviations from the SM couplings will be allowed, one can use an approximate formula for $R_{b}$ obtained by considering only first-order deviations from the tree-level formula
\eq{
R_{b}^{0} = \frac{c_{Lb}^{2} + c_{Rb}^{2}}{\sum_{q}(c_{Lq}^{2} + c_{Rq}^{2})},
}
\eq{
R_{b} = R_{b}^{SM}(1 + 0.5118\,\delta c_{Ld} + 0.5118\,\delta c_{Ls} - 1.8178\,\delta c_{Lb})\,}
where $R_{b}^{SM}$ is given in Table \ref{tab:obs}, and  $\delta c_{Lq}$ are defined in eq. \eqref{eq:clq}.

\subsection{$B_q\to \mu^+\mu^-$}

The rare decays $B_q\to \mu^+\mu^-$, $q=d,s$, are extremely suppressed within the SM because it is mediated by a FCNC and are helicity suppressed. Because of the purely leptonic final state, it is also very clean theoretically and then sensitive to new physics.

The experimental value for $B_s$ is given in table~\ref{tab:obs} and it is of the order of $10^{-9}$.
The branching ratio in the presence of a VLQ can be found in the literature\,\cite{shimizu} and yields
\eq{
Br(B_q\to \mu^+\mu^-) = \tau_{B_{q}} \frac{G_{F}^{2}}{16\pi}\left(\frac{\alpha_{e}}{\pi s^{2}_{w}}\right)^{2}f_{B_{q}}^{2}m_{B^0_{q}}m_{\mu}^{2}\sqrt{1-\frac{4m_{\mu}^{2}}{m^{2}_{B_{q}}}}\,|\eta_{Y}^{2}|\left|\lambda^{t}_{qb}Y_{0}(x_{t})+\Delta^{B}_{\mu\mu}\right|^{2}
\,,
}
where the deviation of the SM is given by
\eq{
\Delta^{B}_{\mu\mu}=-\frac{\pi s^{2}_{w}}{\alpha_{e}}X_{qb}
\,.
}
The factor $\eta_Y=1.0113$ accounts for QCD corrections in a scheme where NLO electroweak corrections can be neglected\,\cite{buras:Bmumu}.

The measurement on $Br(B^0_d\to \mu^+\mu^-)$ still has a large error and will not be considered in the fit. We will instead present predictions for its value based on the fit.
Analogously, the constraint coming from $Br(B^0_s\to X_s\gamma)$ will not be considered because it is weaker than $Br(B^0_s\to \mu^+\mu^-)$\,\cite{shimizu}.

\subsection{$B^0_d\to J/\psi K_S$}

The cleanest measurement of the angle $\beta$ of the $db$ unitarity triangle is given by the $\bar{b}\to c\bar{c}\bar{s}$ transition in $B^0_d\to J/\psi K_S$. It is dominated by only one weak phase in the dominant tree level decay and the penguin pollution is also dominated by the same weak phase.
In the presence of the VLQ, additional contributions are induced.

The measurement of the direct CP violation decay amplitude $S_{\psi K_S}$ occurs through the time-dependent CP asymmetry of $B^0_d$ and $\bar{B}^0_d$ to the common final state $J/\psi K_S$, which is approximately a CP odd state, and can be written as\,\cite{branco:book}
\eq{
\label{Bd->psiKS}
S_{\psi K_S}=\sin(2\beta +2\theta_{B_d}-2\theta_K),
}
where $\beta$ is the angle 
\eq{
\beta\equiv\arg\left(-\frac{V_{cd}V_{cb}^*}{V_{td}V_{tb}^*}\right)
=-\arg\left(-\frac{\lambda^c_{db}}{\lambda^t_{db}}\right)\,,
} 
while $\theta_B$ and $\theta_K$ parametrize the possible deviations from the SM of the mixings in the $B$ and $K$ system as
\eq{
2\theta_{B_d}\equiv\arg\frac{M^{B_d}_{12}}{(M^{B_d}_{12})_{\rm SM}}\,,\quad
2\theta_K\equiv\arg\frac{M^K_{12}}{(M^K_{12})_{\rm SM}}\,.
}
The small phase $\eps'\equiv\arg(-V_{us}V_{cd}V_{ud}^*V_{cs}^*)$\,\cite{aleksan}, contributing as $-2\eps'$ inside the sine in \eqref{Bd->psiKS}, has been neglected.

%%%%%%%%%%%%%%%%%%%%%%%%%%%%%%%%%%%%%%%
\subsection{$K$ meson system: $\eps_K$}
\label{sec:epsK}

The $K$ meson system was the first to show measurable CP violation in the SM.
The value of $|\eps_K|\sim 10^{-3}$ quantifies this indirect CP violation.
The short-distance contribution for the mixing amplitude can be calculated in the presence of a VLQ as (see Ref.\,\cite{saavedra:flavor} and references therein) 
\eq{
M_{12}^{K} = \frac{G_{F}^{2}m_{W}^{2}f^{2}_{K}\hat{B}_{K}m_{K^{0}}}{12\pi^{2}}\left[(\lambda^c_{ds})^2\eta^K_{cc}S_0(x_c)+(\lambda^t_{ds})^2\eta^K_{tt}S_0(x_t)+2\lambda^c_{ds}\lambda^t_{ds}\eta^K_{ct}S_0(x_c,x_t)+\Delta_{K}\right]\,,
}
where
\eq{
\Delta_{K}=-8X_{ds}[\lambda^c_{ds}\eta^K_Z Y_0(x_c)+\lambda^t_{ds}\eta^K_{tt} Y_0(x_t)]+\frac{4\pi s^2_w}{\alpha_e}\eta^K_Z X^2_{ds}\,;
}
the last term being the additional contribution from the VLQ.
The $\eta^K_{ij}$ are QCD correction factors, $\hat{B}_{K}=0.717(24)$\,\cite{PDG} is the bag parameter.

The contribution to the CP violating parameter $\epsilon_K$ is given by
\eq{
\label{eps.K}
\epsilon_K = \kappa_\eps e^{i\phi_\eps}\frac{\text{Im} M_{12}^{K}}{\sqrt{2}\Delta m_{K}}
\,,
}
where $\kappa_\eps\simeq 0.94(2)$\,\cite{buras:epsK} includes effects of $\Delta s$ operators and corrections of deviation of $\phi_\eps=43.52(5)^\circ$\,\cite{pdg} from $45^\circ$.
We use\,\cite{buras.celis.17} $\eta^{K}_{cc} = 1.87(76)$, $\eta^{K}_{tt}=0.5765(65)$, $\eta^{K}_{ct}=0.496(47)$ for the QCD correction factors at NNLO.
We also use $\eta^K_Z=0.60$\,\cite{saavedra:flavor}.
The theoretical prediction for $\Delta m_K$ has large error in long-distance contributions and will not be considered. 
The experimental value is considered in \eqref{eps.K} instead.

%%%%%%%%%%%%%%%%%%%%%%%%%%%%%%%%%%%%%%%
\subsection{$K_L\to\mu\mu$}

A strong limit on $\lambda^t_{sd}$ and $X_{sd}$ comes from the rare decay $K_L\to \mu^+\mu^-$.
Although clean theoretically, the branching ratio $Br(K_L\to \mu^+\mu^-)\sim 10^{-8}$, containing contributions from a dispersive part, $\re^2(A)$, and a absorptive part, $\im^2(A)$, is known to be dominated by the latter, which can be calculated from $Br(K_L\to\gamma\gamma)$.
The theoretical prediction for the dispersive part contains large errors from long distance contributions.
So we consider the simple bound on the short distance (SD) contribution\,\cite{isidori:KLmumu}
\eq{
Br(K_L\to\mu^+\mu^-)_{\rm SD}<2.5\times 10^{-9}\,.
}
A more precise constraint would come by considering the quantity $\chi_{SD}$\,\cite{isidori:KLmumu}
but we limit ourselves to the constraint above.

The short distance part can be calculated by relating it to the decay $K^+\to \mu^+\nu$ as\,\cite{saavedra:flavor}:
\eq{
\frac{Br(K_L\to\mu^+\mu^-)_{\rm SD}}{Br(K^+\to \mu^+\nu)}
=\frac{\tau_{K_L}}{\tau_{K^+}}\frac{\alpha_e^2}{\pi^2s_w^4|V_{us}|^2}
\left[Y_{\rm NL}\re(\lambda^c_{sd})+
\eta^Y_tY_0(x_t)\re(\lambda^t_{sd})+\Delta_{K_L}
\right]\,.
}
We use the factor $Y_{\rm NL}=(2.94\pm 0.28)\times 10^{-4}$ at NLO\,\cite{saavedra:flavor,buchalla.99} and the QCD correction $\eta^Y_t=1.012$\,\cite{saavedra:flavor,buchalla.93}.
Other input values can be found in table \ref{tab:input}.
The VLQ contribution is\,\cite{saavedra:flavor} 
\eq{
\Delta_{K_L}=C_{U2Z}\re X_{sd}\,,
}
where $C_{U2Z}=-\pi s^2_w/\alpha$\,\cite{buras:KL}.

%%%%%%%%%%%%%%%%%%%%%%%%%%%%%%%%%%%%%%%
\subsection{$K$ meson system: $\eps'/\eps$}

The value for $\eps'/\eps$ quantifying direct CP violation in the SM was until recently predicted to be significantly below the current experimental world average from NA48\,\cite{eps':na48} and KTeV\,\cite{eps':ktev} collaborations:
\eq{
(\eps'/\eps)_{\rm exp}=(16.6\pm 2.3)\times 10^{-4}\,;
}
see \cite{buras:epsp:19} for a summary of the discrepancy before April 2020.
However, the recent 2020 calculation of Ref.\,\cite{buras:epsp:20} reports
\eq{
(\eps'/\eps)_{\rm SM}^{(9)}=(13.9\pm 5.2)\times 10^{-4}\,,
}
which is well compatible with the experimental value.
This value takes into account recent lattice calculations of hadronic matrix elements\,\cite{epsp:lattice}, isospin breaking effects\,\cite{epsp:pich} which include the effects of the nonet of lowest-lying mesons and NNLO QCD corrections to EW penguin contributions.

The compatibility weakens the possibility for new physics (NP)\,\cite{buras:epsp:19} as the source of deviation but, due to the large theoretical error, there is still room for a large NP contribution, roughly of the order of\,\cite{buras:epsp:20}
\eq{
-4\times 10^{-4}\lesssim \left(\frac{\eps'}{\eps}\right)_{\rm NP}\lesssim  +10\times10^{-4}\,.
}
We take this interval as a $1\sigma$ range in our fits.

The NP contribution coming from the VLQ can be computed from the simplified formula\,\cite{buras.celis.17}
\eq{
\left(\frac{\eps'}{\eps}\right)_{\rm NP}=
P_7\im(C^{sd}_7)\,,
}
where we consider only the dominant contribution coming from the operator $Q_7$.
The Wilson coefficient at the electroweak scale
\eq{
C^{sd}_7=-\frac{\alpha}{6}\frac{\pi}{G_F^2m_W^2M_B^2}\frac{\tY^B_s{\tY^{B^*}_d}}{\lambda^u_{sd}}\,,
}
depends on the Yukawa coefficients defined in eq.\,\eqref{def:YB-tilde}.
The coefficient in front is
\eq{
P_7=-102.02-1.32 B_6^{(1/2)}+2040.38 B_8^{(3/2)}\,,
}
where we use $B_6^{(1/2)}=1.36\pm 0.23$ and $B_8^{(3/2)}=0.79\pm 0.05$\,\cite{buras:epsp:20} without considering the errors.

%%%%%%%%%%%%%%%%%%%%%%%%%%%
\section{Numerical results}
\label{sec:num}

%%%%%%%%%%%%%%%%%%%%%%%%%%%
\subsection{Methods}

Here we describe the fit procedure.
We use the same set of observables to perform the fit in three cases: (a) SM, (b) addition of one NB-VLQ and (c) addition of one generic VLQ. Therefore, these three cases can be compared directly.
The fit is performed for $M_B=1.4\,\unit{TeV}$, as allowed by direct search constraints \eqref{MB:limit}.

The observables considered in the fit are listed in table \ref{tab:obs}.
For most of the observables, we assume gaussian likelihood and consider
\eq{
\label{chi2:sigma}
\chi^2=-2\ln\mathcal{L}=\sum_{i}\frac{\left(\cO_{i\,\rm theo}-\cO_{i\,\rm exp}\right)^2}{\sigma_i^2}\,, 
}
where $\cO_{i\,\rm theo}$ is the functional form predicted by the model, $\cO_{i\,\rm exp}$ is the central experimental value with error $\sigma_i$.
The exception is the observable $Br(K_L\to\mu^+\mu^-)$, listed in the end of table, which is treated differently from \eqref{chi2:sigma}.
We include it through the function
\eq{
f(x;a,b,r)=
\begin{cases}
\left(\frac{a-x}{r(b-a)}\right)^2, & x\le a,\cr
0, & a<x<b\cr
\left(\frac{x-b}{r(b-a)}\right)^2, & x\ge b,\cr
\end{cases}\,,
}
which is flat in the range $x\in [a,b]$ but grows quadratically outside that range.
The rate of growth is controlled by $r$, chosen to be $r=0.01$.
For $x$ being $Br(K_L\to\mu^+\mu^-)$, this function is added to the $\chi^2$ in \eqref{chi2:sigma} with the ranges $[a,b]$ listed in table \ref{tab:obs}.
The reason for the different treatment of this observable is that it is still subjected to large theoretical uncertainties but, nevertheless, strongly constrain certain BSM parameters. Therefore, we choose a function which does not contribute to the $\chi^2$ if the observable lies within the allowed range but quickly disfavors values which are outside that range.

Regarding the absolute values of the CKM elements in \eqref{|V|:exp}, we do not consider the experimental values for $|V_{td}|$ and $|V_{ts}|$, and instead we use the measured values for $\Delta m_{B_q}$, $q=d,s$, which receive contributions from the VLQ. We also symmetrize the errors and inflate the $1\sigma$ ranges of \eqref{|V|:exp} by 50\% because the high precision of the measurements of the first row of the CKM matrix hints to unitarity violation if taken at face value\,\cite{ramsey:Vud,berezhiani.19,cheung.20}.

The input parameters are listed in table \ref{tab:input}. The parameters $f_{B_d}{\hat B_{B_d}}^{\nhalf}$, $f_{B_s}{\hat B_{B_s}}^{\nhalf}$, $f_{B_s}$ listed in this table, and $\eta^K_{cc}$ described in Sec.\,\ref{sec:epsK}, have large theoretical errors and thus are considered as nuisance parameters which are also included in the fit as in \eqref{chi2:sigma}. We also consider the correlation between the first two of these nuisance parameters.

%%%%%%%%%%%%%%%%%%%%%
\begin{table}
\eq{\nonumber
\begin{array}{|c|c|c|c|}
\hline
\text{Observable} ~~\cO_i & \text{Experimental} & \text{SM (c.v [95\%] CL)} 
\\ \hline 
|V_{ij}| & \text{Eq.\,}\eqref{|V|:exp}^a
        & \text{--} 
\\ \hline 
\Delta m_{B_d}\;[\unit{ps^{-1}}] & 0.5065(19)\,\text{\cite{PDG}} & 0.5060\;[0.5031, 0.5104]
\\ \hline
\Delta m_{B_s}\;[\unit{ps^{-1}}] & 17.749(21)\,\text{\cite{PDG}} & 17.747\;[17.711, 17.787]
\\ \hline 
Br(B_s\to \mu^+\mu^-) 
    & 3.0(4)\times 10^{-9}\text{\cite{PDG}} & (3.40\;[3.13, 3.67]) \times 10^{-9} 
\\ \hline 
|\eps_K| & 2.228(11)\times 10^{-3}\text{\cite{PDG}} & (2.227\;[2.206, 2.248])\times 10^{-3}
\\ \hline 
S_{\psi K_S}~[\sin2\beta_d] & 0.699(17)\text{\cite{PDG}} & 0.71\; [0.68, 0.74]
\\ \hline 
R_b & 0.21629(66)\text{\cite{Rb:Exp}} &  0.21582(11) \text{\cite{Gfitter:Rb}}
\\ \hline 
(\eps'/\eps)_{\rm NP} & [-4,10]\times 10^{-4}\text{\cite{buras:epsp:20}} & \text{--}
\\ \hline\hline
Br(K_L\to \mu^+\mu^-)_{\rm SD} & <2.5\times 10^{-9}\text{\cite{isidori:KLmumu}} & 
(0.86\; [0.76, 0.95]) \times 10^{-9}
\\ \hline
\end{array}
}
\caption{\label{tab:obs}
Observables considered in the fit.
$^a$ We exclude $|V_{td}|,|V_{ts}|$, symmetrize the errors and inflate them in 50\%; see text.
The last column is taken from our fit, except for $R_b$, which is not considered in the SM fit.
}
\end{table}

%%%%%%%%%%%%%%%%%%%%%
\begin{table}
\eq{\nonumber
\begin{array}{|c|c|}
\hline
G_{F}\;[\unit{GeV^{-2}}]& 1.1663787(6)\times 10^{-5}\,\text{\cite{PDG}} 
\\ \hline
s_w^2(m_{Z})~(\overline{\text{MS}})& 0.23121(4)\,\text{\cite{PDG}} 
\\ \hline
m_{W}\;[\unit{GeV}]& 80.379(12)\,\text{\cite{PDG}}  
\\ \hline
m_{Z}\;[\unit{GeV}]& 91.1876(21)\,\text{\cite{PDG}}  
\\ \hline
\alpha(m_{Z})^{-1}~{(\overline{\text{MS}})}& 127.952(9)\,\text{\cite{PDG}}  
\\ \hline
\alpha_{s}(m_{Z})& 0.1179(10)\,\text{\cite{PDG}}  
\\ \hline
m_t(m_t)\;[\unit{GeV}] & 161.9\,\text{\cite{marquard}}^a
\\ \hline
m_c(m_c)\;[\unit{GeV}] & 1.27\, \text{\cite{PDG}}
\\ \hline
y_d(m_{Z}) & 1.58\times 10^{-5}\,\text{\cite{antusch}}
\\ \hline
y_s(m_{Z}) & 3.13\times 10^{-4}\,\text{\cite{antusch}} 
\\ \hline
y_b(m_{Z}) & 1.639\times 10^{-2}\,\text{\cite{antusch}} 
\\ \hline
\hline
\multicolumn{2}{|c|}{\smash{\raisebox{.1\normalbaselineskip}{\text{SM fit}}}}
\\[-1ex] \hline
A & 0.820^{+0.012}_{-0.014}  \\ \hline
\lambda & 0.22655^{+0.00060}_{-0.00066}\\ \hline
10^5\times J & 3.18^{+0.10}_{-0.12}\\ \hline
\end{array}
\begin{array}{|c|c|}
\hline
m_{B_d}\;[\unit{MeV}]& 5279.58(17)\,\text{\cite{PDG}} 
\\ \hline
m_{B_s}\;[\unit{MeV}] & 5366.88(16)\,\text{\cite{PDG}}
\\ \hline
f_{B_d}{\hat B_{B_d}}^{\nhalf} \;[\unit{MeV}]& 225(9)\,\text{\cite{PDG}} 
\\ \hline
f_{B_s}{\hat B_{B_s}}^{\nhalf} \;[\unit{MeV}]& 274(8)\,\text{\cite{PDG}} 
\\ \hline
\rho(f_{B_d}{\hat B_{B_d}}^{\nhalf},f_{B_s}{\hat B_{B_s}}^{\nhalf}) & 
0.951\,\text{\cite{corr:fBds,buras.celis.17}}
\\ \hline
\tau_{B_s} \;[\unit{ps}]& 1.515(4)\,\text{\cite{PDG}} 
\\ \hline
f_{B_s} \;[\unit{MeV}]& 228.4(3.7)\,\text{\cite{flag.19}} 
\\ \hline
m_{K^0}\;[\unit{MeV}]& 497.611(13)\,\text{\cite{PDG}} 
\\ \hline
\Delta m_{K}\;[\unit{MeV}]& 3.484(6)\times 10^{-12}\,\text{\cite{PDG}} 
\\ \hline
f_{\pi}\;[\unit{MeV}]& 130.2(1.2)\,\text{\cite{PDG}} 
\\ \hline
f_K/f_{\pi}\;& 1.192(2)\,\text{\cite{flag.19}} 
\\ \hline
\hline
\multicolumn{2}{|c|}{\smash{\raisebox{.1\normalbaselineskip}{\text{SM fit}}}}
\\[-1ex] \hline
\overline{\rho} & 0.154^{+0.009}_{-0.012} \\ \hline
\overline{\eta} & 0.351^{+0.010}_{-0.008} \\ \hline
&\\ \hline
\end{array}
}
\caption{\label{tab:input}
Input parameters. Some parameters for specific observables are listed in the text.
$^a$ We consider QCD corrections up to four loops with $m_t^{\rm pole}=172.4(7)\,\unit{GeV}$\,\cite{PDG} and $\alpha_s(m_t^{\rm pole})=0.1077$.
The last two rows list the Wolfenstein parameters from the SM fit.
}
\end{table}

To perform the minimization of the $\chi^2$ function and sampling the nearby region we use \multi\,\cite{multinest} accessed through the \textsc{Python} wrapper \pymulti\,\cite{py.multi}.

The Wolfenstein parameters obtained from the SM fit are also listed in table \ref{tab:input}.
We can see they are similar to the fit from PDG in \eqref{pdg:Alamb}, although the values for $A,\bar{\rho}$ and $J$ are larger.

The 95\% CL interval for the observables considered in the fit are also shown in table~\ref{tab:obs}.
We can see that the agreement with the experimental values are excellent.
Moreover, our result for $B_s^0\to\mu^+\mu^-$ is compatible with $Br(B_s^0\to\mu^+\mu^-)=3.23(27)\times 10^{-9}$ of Ref.\,\cite{Bs->mumu:theo} as well as with $Br(B_s^0\to\mu^+\mu^-)=(3.41\;[3.01,3.81])\times 10^{-9}$ from the fit in Ref.\,\cite{buras.celis.17}.
Our intervals for $\Delta m_{B_d}$ and $\Delta m_{B_s}$ are also compatible with Refs.\,\cite{buras.celis.17,deltaM}.
The value of $\eps_K$ agrees with \cite{eps.K}.

%%%%%%%%%%%%%%%%%%%%%%%%%%%
\subsection{Results}

\begin{table}
\eq{\nonumber
\begin{array}{|c|c|c|c|c|c|}
\hline
\text{Observable} ~~\cO_i & \text{Experimental} & \text{SM (c.v [95\% CL])} 
    &\text{NB-VLQ} & \text{gen-VLQ }
\\ \hline
\rule{0pt}{2em} |V_{tb}| & \parbox{4em}{1.013(30)\\[-1ex] \text{[eq.\,\eqref{|V|:exp}]}} 
%     &\parbox{9.3em}{0.999099 \\[-1ex] [0.999054, 0.999156]}
    &\parbox{9.3em}{0.99910 \\[-1ex] [0.99905, 0.99916]}    
    & [0.9982, 0.9992]
    &  [0.9981, 0.9991]
\\ \hline 
10^{10}{\times} Br(B_d^0\to \mu^+\mu^-) {}^a & 1.1^{+1.4}_{-1.3}~\text{\cite{PDG}}
	& 1.08\; [0.99, 1.16]
	&  [0.75, 2.15]
	& [0.0, 8.9]
\\ \hline
10^{10}{\times}Br(K_L\to \mu^+\mu^-)_{\rm SD} & <25~\text{\cite{isidori:KLmumu}}  
    &8.6\; [7.6, 9.5]
    & [4.4, 13.7]
    &[0, 25]
    \\ \hline
10^4{\times}(\eps'/\eps)_{\rm NP} & [-4,10]~\text{\cite{buras:epsp:20}}
    &\text{--}
    & [-6.0, 6.6]
    & [-11, 17] 
\\ \hline 
\end{array}
}
\caption{\label{tab:predicted}
Observables predicted from the fit. The last two columns show the intervals at 95\% CL.
${}^a$ We use the central value for $f_{B_d}=192.0(4.3)$\,\cite{flag.19} without the error.
}
\end{table}
In table \ref{tab:predicted} we show observables that are predicted from the fit.
We extract only the interesting observables and compare the case of SM, one NB-VLQ and one generic VLQ.
For all these cases we show the $95\%$ CL interval for $M_B=1.4\,\unit{TeV}$.

We can see that the value for $|V_{tb}|$ can indeed be lowered with the presence of one VLQ\,\cite{saavedra:flavor} but probing it will require a precision which is hard to attain.

For the rare decay $B_d^0\to \mu^+\mu^-$, our value obtained for the SM is very close to the one obtained in Ref.\,\cite{Bs->mumu:theo}, $Br(B_d^0\to\mu^+\mu^-)=1.06(9)\times 10^{-10}$.
In the presence of one generic VLQ, a large interval is still allowed.
For the NB case, the allowed interval is larger than the SM but narrower than the generic case.

For the kaon decay $K_L\to \mu^+\mu^-$, all the values up to the upper limit can be reached for the generic VLQ case but for the NB-VLQ the interval shrinks to a narrower interval. 
Both these intervals are expectedly wider than for the SM.

A similar conclusion also applies to the VLQ contribution to $\eps'/\eps$: a large contribution is still possible for the generic case but a much smaller contribution is possible for the NB case.
In fact, in the NB case, the predicted maximal contribution at 95\% CL is smaller than allowed experimentally.
The obtained intervals differ from Ref.\,\cite{buras.celis.17} because of the updated calculation reported in Ref.\,\cite{buras:epsp:20}.

\begin{figure}[h]
\includegraphics[scale=0.38]{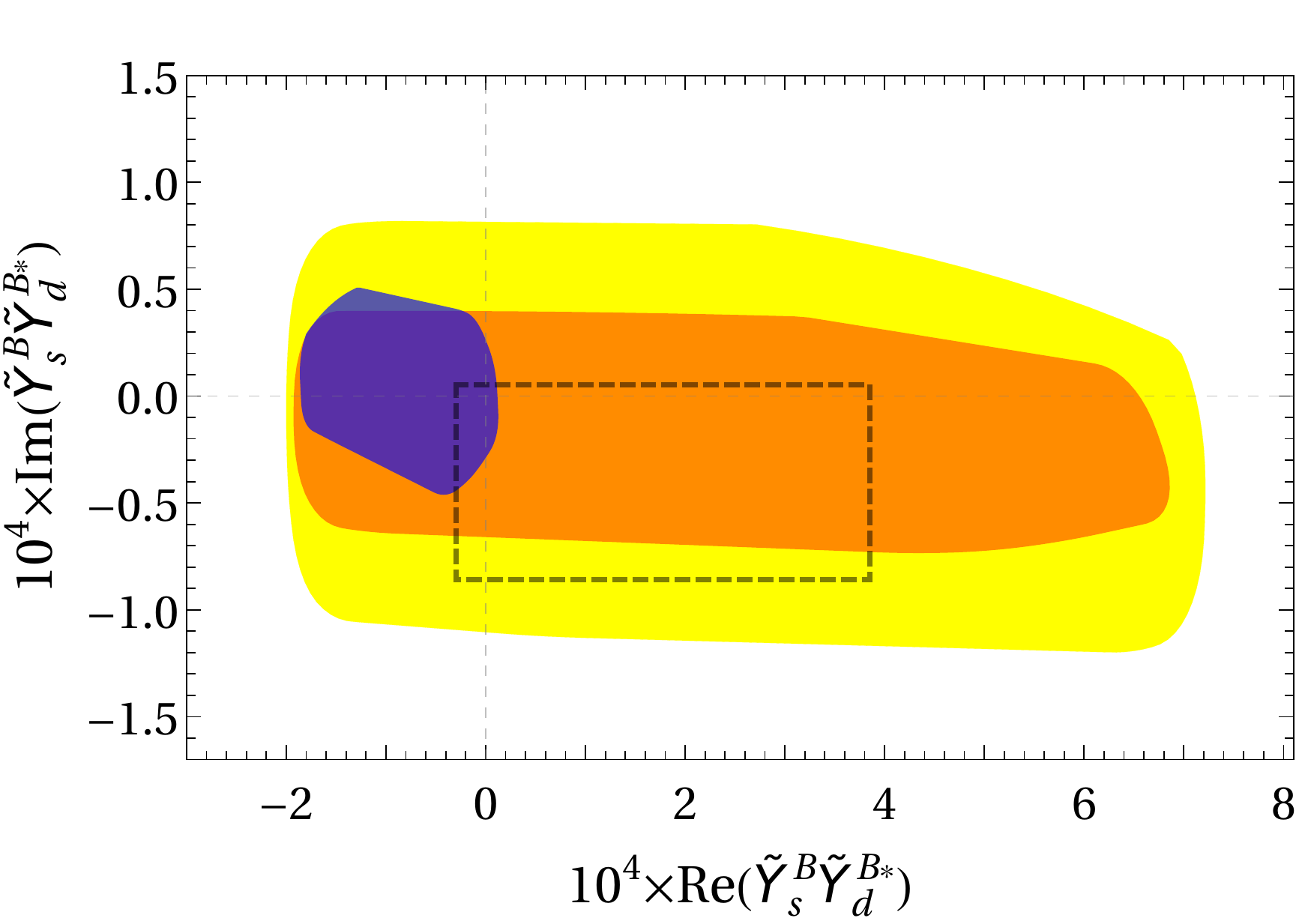}
\caption{\label{fig:Lambda.sd}%
Allowed regions for $\tY^{B}_{s}\tY^{B*}_{d}$ for one generic VLQ (95\% CL in yellow, 68\% CL in orange) 
and one NB-VLQ (95\% CL in blue) for $M_B=1.4\,\unit{TeV}$. The approximate 95\% CL region extracted from the global fit in Ref.\,\cite{buras.celis.17} for $M_B=1\,\unit{TeV}$ is also shown (dashed contour).
}
\end{figure}

\begin{figure}[h]
\includegraphics[scale=0.38]{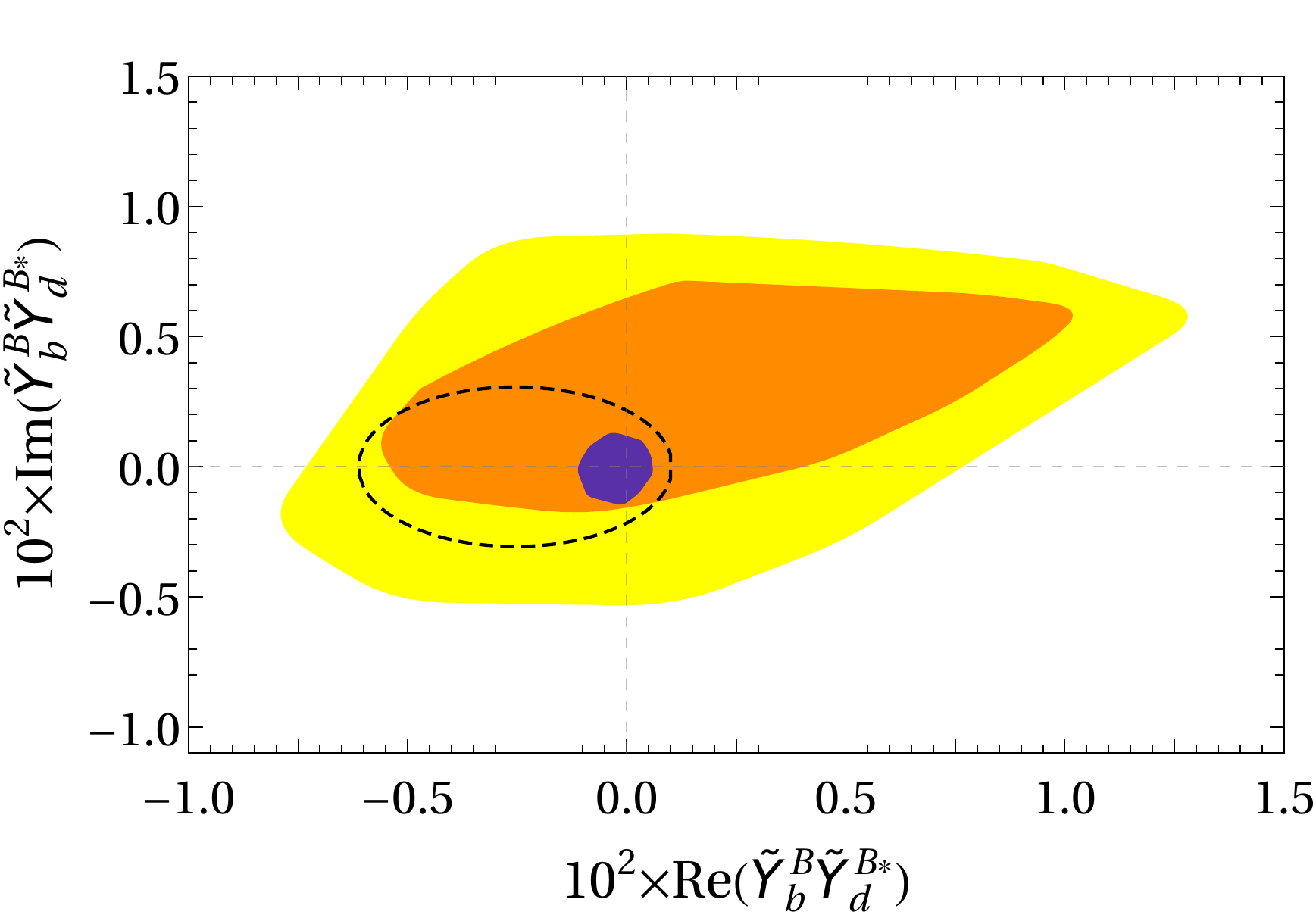}
\caption{\label{fig:Lambda.bd}
Allowed regions for $\tY^{B}_{b}\tY^{B*}_{d}$ for one generic VLQ (95\% CL in yellow, 68\% CL in orange) and one NB-VLQ (95\% CL in blue) for $M_B=1.4\,\unit{TeV}$. The approximate 95\% CL region extracted from the global fit in Ref.\,\cite{buras.celis.17} for $M_B=1\,\unit{TeV}$ is also shown (dashed contour).
}
\end{figure}

\begin{figure}[h]
\includegraphics[scale=0.38]{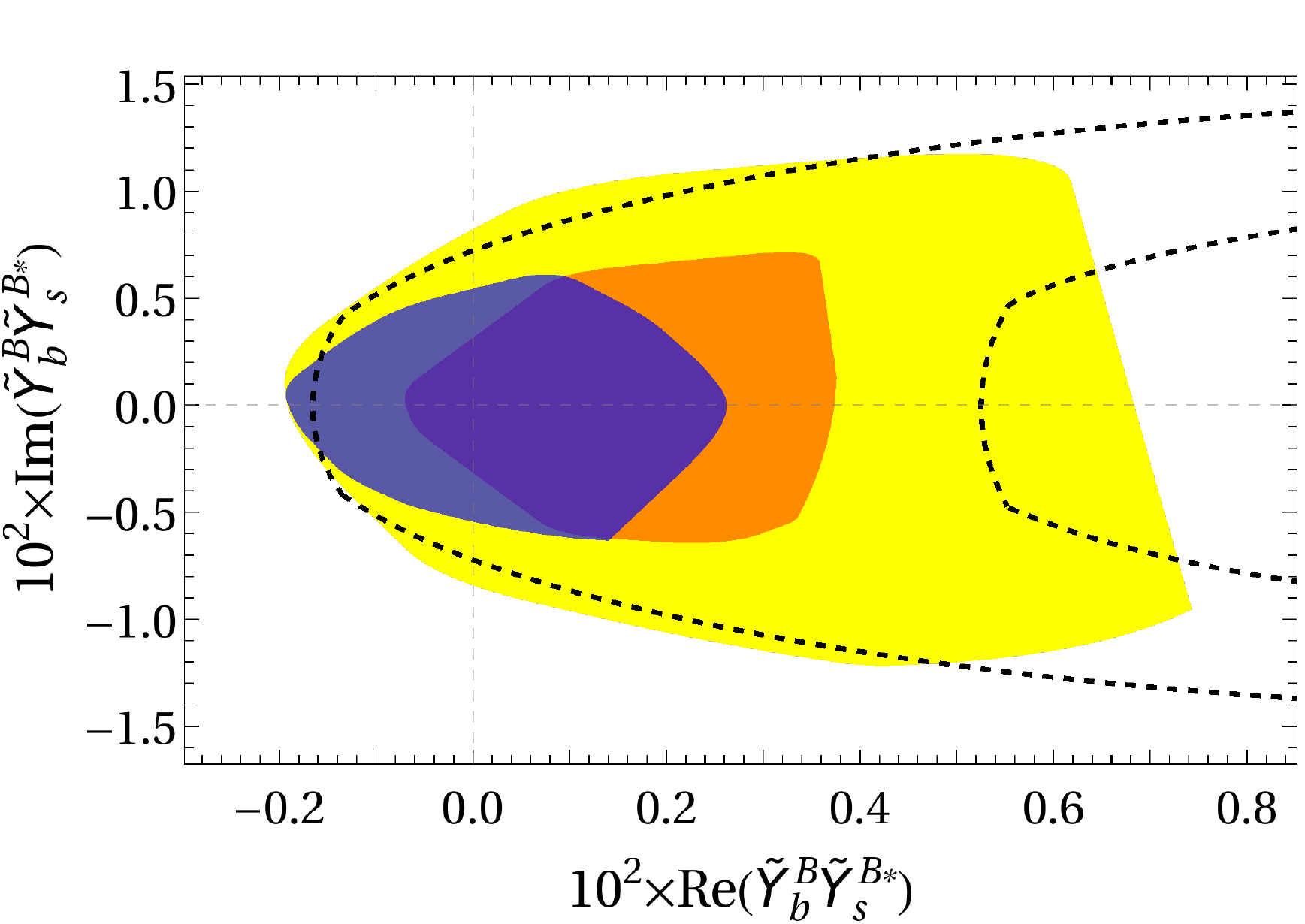}
\caption{\label{fig:Lambda.bs}
Allowed regions for $\tY^{B}_{b}\tY^{B*}_{s}$ for one generic VLQ (95\% CL in yellow, 68\% CL in orange) and one NB-VLQ (95\% CL in blue) for $M_B=1.4\,\unit{TeV}$. The approximate 95\% CL region extracted from the global fit in Ref.\,\cite{buras.celis.17} for $M_B=1\,\unit{TeV}$ is also shown (dashed contour).
}
\end{figure}

We show in Figs.\,\ref{fig:Lambda.sd}, \ref{fig:Lambda.bd} and \ref{fig:Lambda.bs} the allowed regions for the combination $\tY^B_i\tY^{B*}_j$, cf.\,\eqref{def:YB-tilde}, 
in the complex plane for the sectors $(ij)=(sd),(bd),(bs)$, respectively.
These quantities were denoted as $\Lambda_{ij}$ in Ref.\,\cite{buras.celis.17} and can be compared here.
The region in yellow (orange) corresponds to the generic VLQ case at 95\%~CL (68\%~CL) while the region in blue corresponds to the NB-VLQ case at $95\%$~CL. In all sectors the allowed regions for the NB-VLQ case are much smaller than the generic case due to the correlations of the model. In the sectors $(sd)$ and $(bd)$, the 95\% CL region for the NB case is even smaller than the 65\% CL region of the generic case.  
For comparison, we also show in a dashed contour the approximate 95\% CL regions from the global fit in Ref.\,\cite{buras.celis.17} which used $M_B=1\,\unit{TeV}$ (our value is $M_B=\,\unit{1.4}\,\unit{TeV}$).
Some comments are in order.

\begin{itemize}
\item For the $(sd)$ sector, we use the most constraining observables as in Ref.\,\cite{buras.celis.17}:
$Br(K_L\to\mu\bar{\mu})_{\rm SD}$ and $(\eps'/\eps)_{\rm NP}$.
The region is constrained horizontally and vertically, respectively, by these observables.
Our implementation of $Br(K_L\to\mu\bar{\mu})_{\rm SD}$, however, is different and we obtain a region wider horizontally. 
Vertically, we consider an updated (2020) constraint for $(\eps'/\eps)_{\rm NP}$\,\cite{buras:epsp:20} which differs from the constraint in Ref.\,\cite{buras.celis.17} of 2016.
The net result is an upward shift in the region.
The region is possibly wider due to the change in the VLQ mass from 1 TeV to 1.4 TeV.

\item For the $(bd)$ sector, our region for the generic case is much larger than the region found in Ref.\,\cite{buras.celis.17} because we do not include the observable $Br(B^+\to \pi^+\mu\bar{\mu})$ which is the most constraining for that mass range. Nevertheless, the regions are roughly compatible and our region for NB case is entirely contained.

\item For the $(bs)$ sector, we use the most constraining observables as in Ref.\,\cite{buras.celis.17}: $Br(B_s\to\mu\bar{\mu})$. The obtained region for the generic case is largely compatible, with possible enlargement due to the increase in $M_B$.
For the NB case, the region is much narrower in the imaginary part.

\end{itemize}

Note that, after we factor $V_{d_L}$, the mixing $\Theta_i$ in \eqref{V} of up-quarks with the heavy $B$ quark through $W$ exchange is given by
\eq{
\Theta_i= \frac{v}{\sqrt{2}}\tY^B_{i}/M_B,
}
i.e., it is proportional to $\tY^B_i$ in \eqref{def:YB-tilde}; see eqs.\,\eqref{Theta} and \eqref{thetaL}.
So the constraints in Figs.\,\ref{fig:Lambda.sd}, \ref{fig:Lambda.bd} and \ref{fig:Lambda.bs} can be interpreted as constraints on $\Theta_i\Theta_j^*$.
Similarly, flavor changing neutral currents couple to the $Z$ boson through $X^d$ in \eqref{X:d}, which is proportional to 
\eq{
X^d_{ij}\approx -\Theta_i\Theta_j^*= -\frac{v^2}{2M_B^2}(\tY^B_{i}\tY^{B*}_{j})\,,\quad
i\neq j\,.
}
The approximation refers to the leading seesaw approximation.
The factor inside the parentheses is exactly the quantity appearing in Figs.\,\ref{fig:Lambda.sd}, \ref{fig:Lambda.bd} and \ref{fig:Lambda.bs}.
Indeed, if we link the flavor changing physics scale in \eqref{Lambda.scale} to $1/\Lambda_{ij}^2\sim |\tY^B_{i}\tY^{B*}_{j}|/M_B^2$, apart from a global factor, we can extract from the figures the milder hierarchies for the NB case,
\eq{
\Lambda_{sd}^{-1}:\Lambda_{bd}^{-1}:\Lambda_{bs}^{-1}\sim 
0.18:0.4:1\,.
}
In part, this is due to a larger set of observables we consider, which is not restricted to $\Delta F=2$ observables.
It is important to emphasize that the NB scheme leads to a \emph{hierarchy} of $\tY^B_i$ inherited from the SM Yukawa couplings but the overall scale is not determined\,\cite{nb-vlq}.
In special, such a scale is not suppressed by the bottom Yukawa, such as in the MFV setting, and is only limited by perturbativity from the theoretical side while flavor observables in the $(bs)$ sector, cf.\;Fig.\,\ref{fig:Lambda.bs}, allows $|\tY^B_b|\sim 0.3$ for $M_B=1.4\,\unit{TeV}$, considering that $|\tY^B_b\tY^{B*}_s|\sim |V_{ts}(\tY^B_b)^2|$.
For larger $M_B$, larger values are allowed.
See appendix \ref{ap:approx} for approximate formulas for $\tilde{Y}^B$.

\begin{figure}[h]
\includegraphics[scale=0.4]{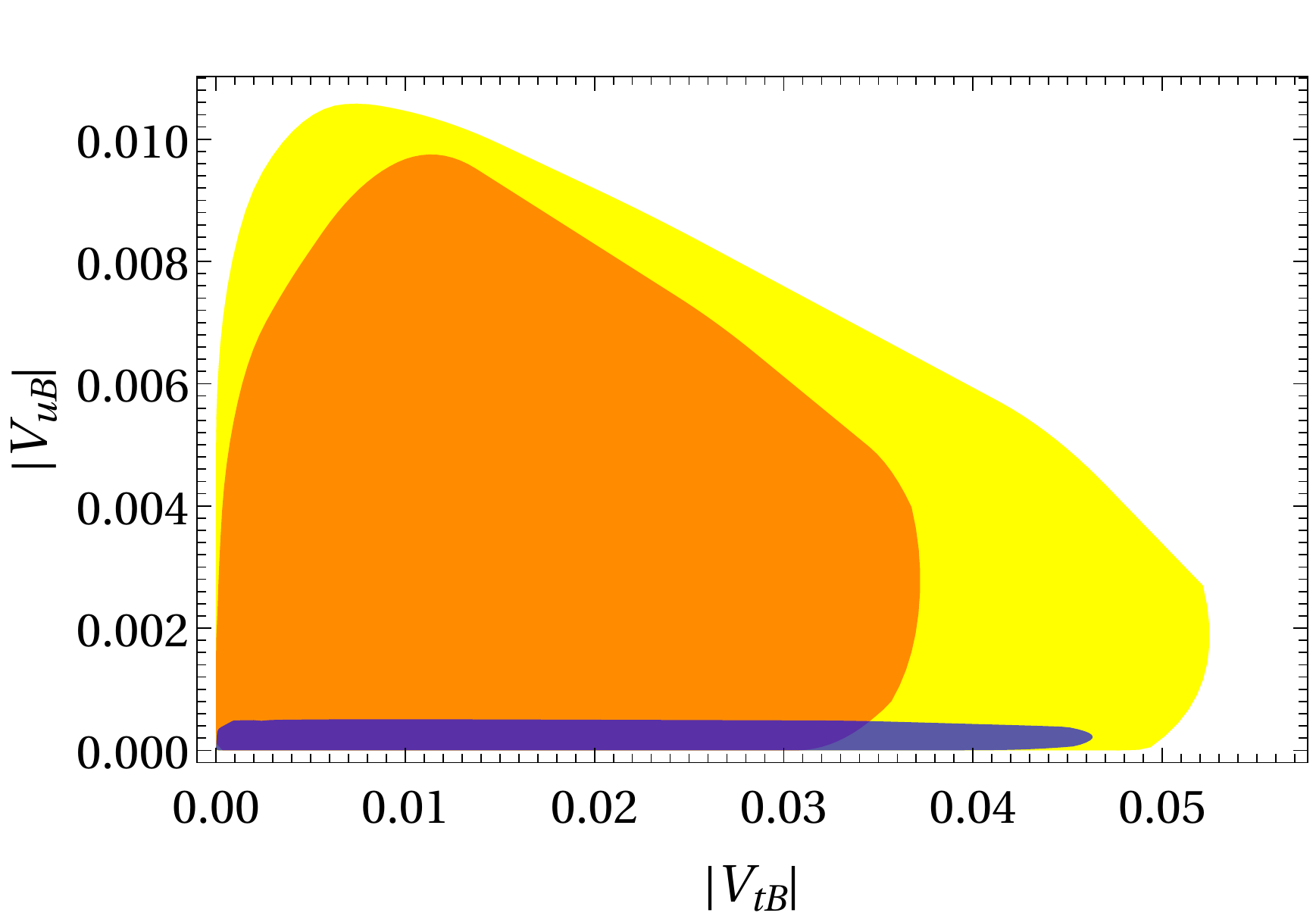}
\includegraphics[scale=0.4]{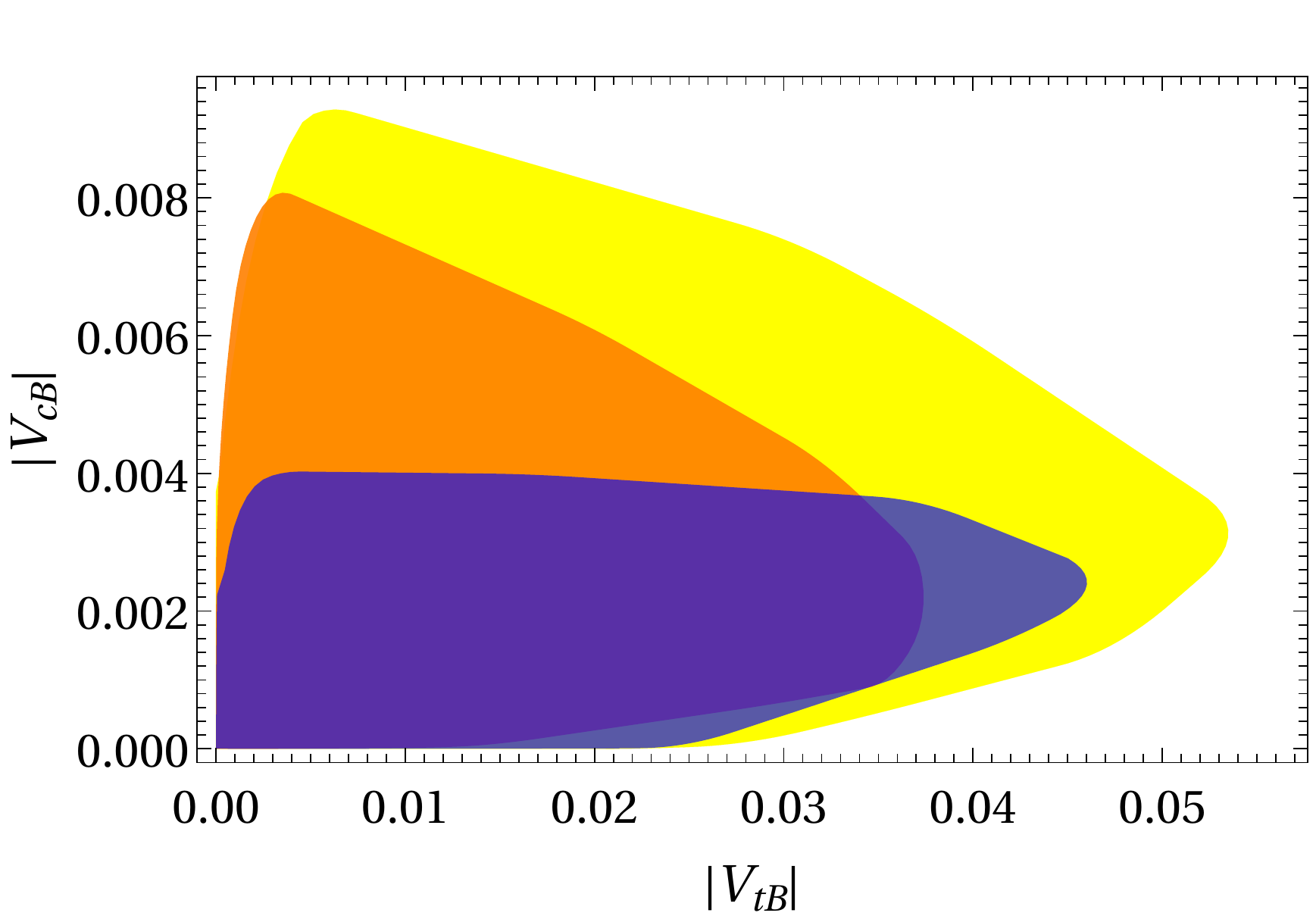}
\caption{\label{fig:ViB}%
Allowed regions of $|V_{iB}|$ for one generic VLQ (95\% in yellow, 68\% in orange) and one NB-VLQ (95\% CL in blue).
}
\end{figure}

There is one frequently considered simplified scenario where the combinations appearing in Figs.\,\ref{fig:Lambda.sd}, \ref{fig:Lambda.bd} and \ref{fig:Lambda.bs} all vanish and no constraint can be extracted.
That is the case where the heavy quark $B$ only couples with the third SM family\,\cite{saavedra:handbook} and 
\eq{
\tY^B=(0,0,y_B)^\tp\,.
}
Similar conclusions apply to scenarios where $B$ couples only to the first family or only to the second family.
For these cases, it is more interesting to show directly the constraint on $|V_{iB}|\approx |(V_{d_L}\Theta)_i|$, which also quantifies the deviation from unitarity for the $3\times 3$ sub-block.
We show the allowed regions in Fig.\,\ref{fig:ViB}.
The color code is the same as in Fig.\,\ref{fig:Lambda.sd}.
We can see that for the generic case (yellow/orange) all three $|V_{iB}|$ can reach values of the same order of magnitude (not simultaneously) but for the NB case (blue), the allowed values are hierarchical for $i=t,c,u$.
The latter pattern is compatible with the hierarchy \eqref{ViB:hierarchy} but it follows \eqref{ViB:3f} only roughly. 
Therefore, it is clear that $|V_{tB}|$ (similarly $|Y^B_{3}|$) can reach values as large as in the generic case but the other mixings (Yukawas) are automatically suppressed.

The constraint for the mixing $|V_{iB}|$ has another advantage: it is more applicable to different VLQ masses. The reason is that our flavor observables are essentially sensitive to $X^d$ which depends on the CKM matrix $V$. The mixing $V_{iB}$, however, scales approximately as \eqref{thetaL} and a larger mass implies a smaller mixing for the same Yukawa $Y^B$.
So the constraints in Figs.\,\ref{fig:Lambda.sd}, \ref{fig:Lambda.bd} and \ref{fig:Lambda.bs} can be approximately adapted to other VLQ masses after rescaling while the constraints on Fig.\,\ref{fig:ViB}
can be approximately applied to any mass.\,\footnote{%
This scaling breaks down for masses larger than tens of TeV due to the one-loop box contribution that starts to dominate the mixings of neutral mesons\,\cite{ligeti.wise}.}

\begin{figure}[h]
\includegraphics[scale=0.5]{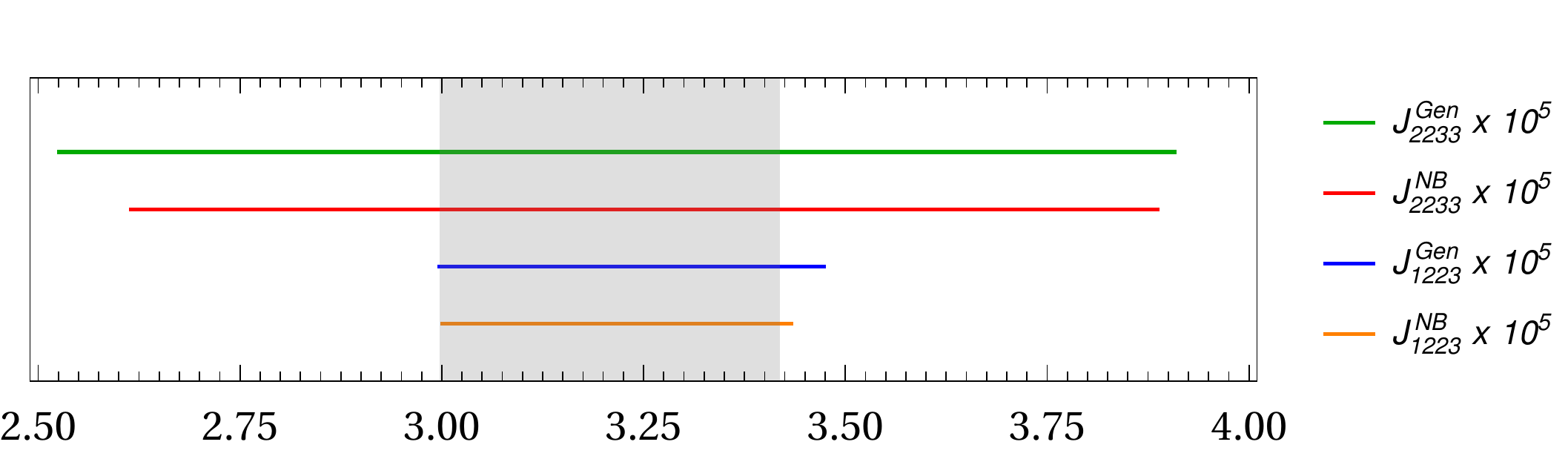}
\caption{\label{fig:J95}%
95\% CL intervals for Jarlskog invariants, cf.\,\eqref{Jijkl}, for the case of one generic VLQ and one NB-VLQ. The gray band is the 95\% CL interval for the SM.
}
\end{figure}

Concerning CP violation, we show in Fig.\,\ref{fig:J95} the 95\% CL intervals for some Jarlskog invariants involving only the $3\times 3$ sub-block of CKM.
The definition we use is\,\cite{nb-vlq}
\eq{
\label{Jijkl}
J_{ijkl}=\im[V_{ij}V^\dag_{jk}V_{kl}V^\dag_{li}]\,.
}
The usual Jarlskog invariant for the SM is $J_{uscb}=J_{1223}$ and we show in the gray band the allowed 95\% interval obtained from the fit.
With the presence of one VLQ, the CKM matrix becomes $3\times 4$ and the Jarlskog invariants are not all the same.
We also show the intervals for $J_{1223}$ and $J_{2233}$ for a generic VLQ and for a NB-VLQ.
The invariant $J_{2233}$ was shown in Ref.\,\cite{nb-vlq} to present a larger variation for one NB-VLQ.
We can indeed see in Fig.\,\ref{fig:J95} that the interval for $J_{2233}$ with one VLQ is larger than in the SM for both the generic and the NB case.
The deviation of $J_{2233}$ from the SM value we see in Fig.\,\ref{fig:J95} could be tested in the future by a more precise determination of $\phi_s$ proportional to the angle between $V_{cs}V_{cb}^*$ and $V_{ts}V_{tb^*}$ which enters precisely in $J_{2233}=\im[V_{cs}V^*_{ts}V_{tb}V^*_{cb}]$.
Currently, the errors are no smaller than 40\%\,\cite{phi_s} but the precision at LHCb at High-luminosity LHC with 300 fb$^{-1}$ is expected to be around 10\%.\,\cite{LHCb.high}.

Concerning the 5 BSM parameters \eqref{param.bsm} of our seesaw parametrization for one NB-VLQ, 
there are essentially no constraints on the parameters $b,\beta_1,\beta_2,\gamma$.
The mass $M_B$ can be constrained in direct searches in colliders but are not directly constrained by flavor observables which are dependent on the mixing between the SM quarks and the VLQ.

For a qualitative comparison, we briefly discuss the case of one up-type NB-VLQ in appendix \ref{ap:up}.
The salient features are the same: the Yukawa couplings $Y^T_i$ of the heavy up-type VLQ $T_{L,R}$ with the SM quarks are \emph{hierarchical} reflecting the hierarchy in the up-type Yukawas and in the CKM mixing.
For this reason, for $M_T\sim 1.3\,\unit{TeV}$, obeying flavor diagonal electroweak precision constraints automatically suppresses flavor changing observables involving lighter families.

%%%%%%%%%%%%%%%%%%%%%%%%%%%
\section{Conclusions}
\label{sec:concl}

By performing a global fit on interesting low energy flavor observables we constrain the possibilities for the presence of one VLQ of charge $-1/3$.
We compare the case of one generic VLQ and one VLQ of Nelson-Barr type. Only the latter would be responsible for transmitting the spontaneous CP violation from a CP breaking sector to the SM in the Nelson-Barr setting.
We see that, using flavor observables, distinguishing the case of one NB-VLQ from a generic VLQ of down type will be a challenging task.
Nevertheless the presence of one NB-VLQ is definitely falsifiable once a VLQ is found by other means and its flavor signature lies in a region of parameter space outside of what is allowed for one NB-VLQ but within the region allowed for one generic VLQ.
The results presented in table \ref{tab:predicted} and Figs.\,\ref{fig:Lambda.sd}, \ref{fig:Lambda.bd} and \ref{fig:Lambda.bs} show that the parameter space constrained by flavor observables is significantly smaller for the NB case compared to the generic case due to the inherent correlations that appear in the model, a feature which was uncovered in Ref.\,\cite{nb-vlq}.

In the process, we update the allowed parameter space for the presence of one generic VLQ, which can be directly compared to the regions obtained in Ref.\,\cite{buras.celis.17}.
These constraints should apply to all limiting cases where one VLQ of down type is present.
For example, the often considered case of a VLQ coupling only to the third family is just a subcase and must have the allowed parameter space contained in the generic region.
On the other hand, the NB case is a specific example of a scenario where the VLQ couples dominantly with the third family but also couples to the lighter families with hierarchically smaller couplings.

In summary, scenarios involving VLQs of Nelson-Barr type are simple scenarios where the presence of VLQs are motivated to solve the strong CP problem.
Distinguishing a VLQ of NB type from a VLQ unrelated to the origins of CP breaking in nature is a very difficult but interesting task that warrants further investigation.

%%%%%%%%%%%%%%%%%%%%%%%%%%%
\acknowledgements

C.C.N.\ acknowledges partial support by Brazilian Fapesp, grant 2014/19164-6, and
CNPq, grant 304262/2019-6. 
G.D.C.\ acknowledges financial support by the Coordenação de Aperfeiçoamento de Pessoal de Nível Superior - Brasil (CAPES) - Finance Code 001.
We thank Andre Lessa for helpful comments.

%%%%%%%%%%%%%%%%%%%%%%%%%%%%%%%%%%%%%%%%%%%
\appendix
%%%%%%%%%%%%%%%%%%%%%%%%%%%%%%%%%%%%%%%%%%%
\section{Inversion formula}
\label{ap:inversion}

Here we briefly review the inversion formula for \eqref{Yd:NB} developed in Ref.\,\cite{nb-vlq} for the seesaw parametrization of one NB-VLQ.
The Yukawa couplings $Y^d$ in \eqref{Yd:NB} and $Y^B$ in \eqref{YB:NB} depend on the quantities $\cY^d$ and $w$ but $Y^d$ should be in accord with SM input \eqref{Yd:sm.input}.
The inversion formula allows us to take the latter into account automatically, leaving $\cY^d$ and $w$ dependent on $\{b,\gamma,\beta_1,\beta_2\}$. The VLQ mass $M_B$ completes the five free BSM parameters.

Given $\beta_1,\beta_2$ in \eqref{VdL}, $\cY^d$ in \eqref{Yd:NB} can be determined as
\eq{
\label{formula:cal-Yd}
\cY^d=A\cO B^{-1}
}
where 
\eqali{
\label{def:A}
A&= \Big(\re(Y^d{Y^d}^\dag)\Big)^{1/2}\,,
\cr
B&= \diag(1,\sqrt{1-b^2},\sqrt{1-a^2})\,,
}
and the orthogonal matrix $\cO$ will be defined below.
The matrix $B$ contains one free parameter, chosen to be $b$, while $a$ is fixed from
\eq{
\label{def:mu}
\frac{a}{\sqrt{1-a^2}}\frac{b}{\sqrt{1-b^2}}=\mu\,.
}
The value of $0<\mu<1$ comes from the real antisymmetric matrix
\eq{
C\equiv A^{-1}\im(Y^d{Y^d}^\dag)A^{-1}\,,
}
with eigenvalues $(0,i\mu,-i\mu)$. The real orthogonal matrix $\cO$ in the formula \eqref{formula:cal-Yd} transforms $C$ to the canonical form
\eq{
\label{def:cO}
\cO^\tp C\cO=\mu\mtrx{1&&\cr&0&-1\cr&1&0}\,.
}
This matrix $\cO$ is not unique as \eqref{def:cO} is invariant by additional rotation on the 2-3 plane.
Then, we can write
\eq{
\cO=\cO_0\mtrx{1&&\cr &\cos\gamma &\sin\gamma\cr &-\sin\gamma&\cos\gamma}\,,
}
where $\cO_0=(e_1|e_2|e_3)$ is a representative matrix satisfying \eqref{def:cO}.
The first column vector $e_1$ is fixed as the eigenvector of $C$ with eigenvalue zero. 
The sign is determined so that $\cY^d_{11}>0$.
For our numerical code, we choose $e_2$ and $e_3$ so that $u_-=e_2+ie_3$ is the eigenvector of $C$ with eigenvalue $-i\mu$.
We choose the convention that $u_-$ has real positive first component so that $e_3$ has vanishing first component.

As for the parameter ranges, the angular and phase variables, $\gamma,\beta_1,\beta_2$, should all be confined to $[0,2\pi]$. The $b$ parameter is confined to the range \eqref{b.range} so that the quality of the seesaw approximation is guaranteed within 1\%.

%%%%%%%%%%%%%%%%%%%%%%%%%%%%%%%%%%%%%%%%%%%
\section{Approximate formulas for the Yukawas}
\label{ap:approx}

The inversion formula reviewed in appendix \ref{ap:inversion} solves for $\cY^d$ in \eqref{Yd:NB} in terms of the SM Yukawa $Y^d$ and this relation is exact in the leading seesaw approximation.
We can obtain approximate relations for $Y^B$ which makes explicit the hierarchy of these couplings.

Substituting the solution \eqref{formula:cal-Yd} in \eqref{YB:NB} we obtain
\eq{
Y^B=\left[\re(Y^d{Y^d}^{\dag})\right]^{1/2}\cO\mtrx{0 \cr i x_b \cr \dst\frac{\mu}{x_b}}
\,,
}
where $x_b\equiv\frac{b}{\sqrt{1-b^2}}$ and we have used \eqref{def:mu}.
Considering $Y^d$ is hierarchical and ignoring the real part, we obtain for $b\le 1/\sqrt{2}$ the approximate bound
\eq{
\label{YB:approx}
|Y^B_i|\lesssim y_b|V_{ib}|\frac{\mu}{x_b}
\sim \frac{\mu}{x_b}\times (10^{-4},10^{-3},0.024)
\,. 
}
Since $|V_{iB}|\approx |(\theta_L)_i|=\frac{v}{\sqrt{2}}|Y^B_i|/M_B$, this expression describes to a good approximation the upper boundary of Fig.\,3 in Ref.\,\cite{nb-vlq} as a function of $b$ (typically $\mu\sim 1$).
As the CKM matrix is hierarchical, we can also write $|\tilde{Y}^B_i|\sim |Y^B_i|$.
Equation \eqref{YB:approx} clearly shows the hierarchy among the couplings $Y^B_i$. It also shows that \emph{the absolute scale} of the Yukawa couplings of the NB-VLQ are \emph{not} necessarily suppressed by the bottom Yukawa for $x_b\ll 1$. Indeed, our fit showed that it can reach $|Y^B_3|\sim y_b/x_b\sim 0.3$ for $M_B=1.4\,\unit{TeV}$ in conformity with other fits of the \emph{generic} VLQ case\,\cite{buras.celis.17}.

For completeness, we can quantify the quality of the leading seesaw approximation by considering the deviation of the real heaviest mass eigenvalue of \eqref{mass.matrix} compared to the input mass $M_B$:
\eq{
\label{delta.M}
\frac{\delta M_B}{M_B}\approx\ums{2}|\theta_L|^2\lesssim \frac{m_b^2}{2M^2_B}\frac{\mu^2}{x^2_b}\,.
}
The mixing $\theta_L$ is related to $Y^B$ by \eqref{thetaL}.
With this relation, one can limit the maximum deviation allowed by setting a lower limit for $b$.

%%%%%%%%%%%%%%%%%%%%%%%%%%%%%%%%%%%%%%%%%%%
\section{Auxiliary functions}
\label{ap:aux}

The Inami-Lim functions\,\cite{inami.lim} are
\eqali{
S_0(x,y)&=
\frac{x \left(x^2-8 x+4\right) y \log (x)}{4 (x-1)^2 (x-y)}+\frac{x \left(y^2-8 y+4\right) y \log (y)}{4 (y-1)^2 (y-x)}-\frac{3 x y}{4 (x-1)(y-1)}
\cr
S_0(x)&=\frac{x^3-11 x^2+4 x}{4 (1-x)^2}-\frac{3 x^3 \log (x)}{2 (1-x)^3}
\,,
\cr
Y_0(x)&=\frac{3 x^2 \log (x)}{8 (x-1)^2}-\frac{3 x}{8 (x-1)}+\frac{x}{8}\,.
}
We follow the convention of Ref.\,\cite{saavedra:flavor}.
Note that $S_0(x)=\lim_{y\to x}S_0(x,y)$.

%%%%%%%%%%%%%%%%%%%%%%%%%%%%%%%%%%%%%%%%%%%
\section{One up-type NB-VLQ}
\label{ap:up}

In this case, the down-type VLQs $B_{L,R}$ are replaced by up-type VLQs $T_{L,R}$ and the Lagrangians  
\eqref{yuk:NB} (real basis) and \eqref{yuk:VLQ} (generic basis) can be adapted accordingly.
The Yukawa couplings that will be constrained by flavor observables are
\eq{
-\lag\subset \tilde{Y}^T_i\bar{q}_{iL}\tilde{H}T_R\,,
}
in the basis where $Y^u$ is diagonal and $Y^d$ carries the CKM of the SM.
Note that $T$ in $\tilde{Y}^T$ is a label and not the transpose operation. To avoid confusion, we will omit this label in the following discussion.
Similarly, $Y=Y^T$ without the tilde is the same Yukawa in the basis where $Y^d$ is diagonal.
Because of CKM hierarchy, $|\tY_i|\sim |Y_i|$.

The full CKM matrix will be $4\times 3$ of the form
\eq{
V\approx 
\left(\begin{array}{c}
    \id_3-\ums{2}\Theta\Theta^\dag\\
    \hline
    \Theta^\dag 
    \end{array}
\right)
V_{u_L}^\dag
\,,
}
where $V_{u_L}^\dag=V_{ckm}$ is the $3\times 3$ CKM matrix within the SM, except for possible two additional phases.
The coupling $\tY$ controls the mixing of the VLQ with the SM quarks as, in leading order, 
\eq{
\Theta_i=\frac{v}{\sqrt{2}}\tilde{Y}_i M_T^{-1}\,.
}
The FCNC coupling to the $Z$ now depends on $X^u=VV^\dag$.

In order to reproduce the SM up-type Yukawa couplings, a relation analogous to \eqref{Yd:NB} needs to be satisfied and the up-type NB-VLQ will also couple to SM quarks hierarchically.\footnote{%
Note that due to $y_t\gg y_b$, the leading seesaw approximation is much less precise for an up-type NB-VLQ than for a down-type NB-VLQ, cf.\,\eqref{delta.M}. But once we consider \eqref{constraint:Theta3}, the approximation is better than $1\%$.
}
Owing to this hierarchy in the VLQ mixing, we can use the constraints from the case where the VLQ couples only to the third family.
For example, the oblique parameters $S,T$ constrain\,\cite{saavedra:handbook}
\eq{
\label{constraint:Theta3}
|V_{Tb}|\approx |\Theta_3|\lesssim 0.12\,,
}
for $M_T>1.3\,\unit{TeV}$.
This leads to
\eq{
\label{constraint:YT3}
|\tilde{Y}_{3}|\lesssim 0.9\times\left(\frac{M_T}{1.3\,\unit{TeV}}\right)\,.
}
This constraint is stronger than other electroweak observables that deviate due to modified $Zbb$ coupling.

Adapting the approximate relation \eqref{YB:approx} to an up-type NB-VLQ, we obtain
\eq{
\label{YT:approx}
|Y_i|\lesssim y_t|V_{ti}^*|\frac{\mu}{x_b}
\sim \frac{\mu}{x_b}\times (0.009,0.04,1)
\,. 
}
Therefore, if we keep the hierarchy in \eqref{YT:approx},\footnote{%
Strictly speaking, this hierarchy may be subjected to an order of magnitude variation as the Yukawas do not need to saturate the upper limits in \eqref{YT:approx}; see Fig.\,3 in Ref.\,\cite{nb-vlq} for the down-type NB-VLQ. In fact, to respect \eqref{constraint:YT3}, $Y_3$ will be far from saturating the upper limit. 
}
the constraint \eqref{constraint:YT3} naturally translates into a much stronger constraint on the other Yukawas:
\eq{
\label{YT:12}
|\tilde{Y}_{1}|\lesssim 0.008\times\left(\frac{M_T}{1.3\,\unit{TeV}}\right)\,,\quad
|\tilde{Y}_{2}|\lesssim 0.04\times\left(\frac{M_T}{1.3\,\unit{TeV}}\right)\,.
}
So due to the hierarchy of $\tilde{Y}$, analogously to the down-type NB-VLQ, the Yukawa couplings of the up-type NB-VLQ with lighter quarks are automatically suppressed.
As a result, flavor changing observables involving lighter quarks may not be as relevant as for a generic up-type VLQ without the hierarchy in the couplings.

Ref.\,\cite{ligeti.wise} analyzed various flavor observables for all VLQ representations coupling to the SM. 
Due to the hierarchy in \eqref{YT:approx}, the constraint \eqref{constraint:YT3} coming from flavor diagonal electroweak precision observables is more important than flavor violating ones.
For example, the decay $D^0\to\mu^+\mu^-$ constrains the combination\,\cite{ligeti.wise}
\eq{
|\tilde{Y}_1\tilde{Y}_2^*|<0.11\times\left(\frac{M_T}{1.3\,\unit{TeV}}\right)^2
\,.
}
But the combination of the limits in \eqref{YT:12} is much stronger.
Analogous considerations apply to other limits.

Although not very important for $M_T\sim\unit{TeV}$, the one-loop contributions to neutral meson mixings may become important for larger masses because the amplitudes scale as $Y^4/M_T^2$ (box diagram) instead of $Y^2/M_T^2$.
Let us collect the results of Ref.\,\cite{ligeti.wise}:
\eqali{
|\im Y_1 Y^*_2|&<0.002\times \frac{M_T}{1.3\,\unit{TeV}}\,,
\cr
|\re Y_1Y^*_2|&<0.03\times \frac{M_T}{1.3\,\unit{TeV}}\,,
\cr
|Y_1Y^*_3|&<0.05\times \frac{M_T}{1.3\,\unit{TeV}}\,,
\cr
|Y_2 Y^*_3|&<0.2\times \frac{M_T}{1.3\,\unit{TeV}}\,.
}
Notice the different scaling in $M_T$.
The last two constraints become more important than \eqref{constraint:YT3}, combined with \eqref{YT:approx}, for $M_T\sim 7\,\unit{TeV}$.

%%%%%%%%%%%%%%%%%%%%%%%%%%%%%%%%%%%%%%%%%%%

%%%%%%%%%%%%%%%%%%%%%%%%%%%%%%%%%%%%%%%%%%%%%%%%%
\end{document}